\documentclass[conference]{IEEEtran}
\IEEEoverridecommandlockouts
\usepackage[table,xcdraw]{xcolor}
\usepackage{mathtools}
\usepackage{amssymb}
\usepackage{algorithm,algpseudocode}
\usepackage[binary-units]{siunitx}
\usepackage{flexisym}
\usepackage{color,soul}
\usepackage{textgreek}
\usepackage{multirow}
\usepackage{multicol}
\usepackage[export]{adjustbox}
\usepackage{footnote}
\usepackage{makecell}
\usepackage{url}
\usepackage{comment}
\usepackage{pgfplots,filecontents}
\usepackage{tikz}
\usepackage{subcaption}
\usepackage{graphicx}
\usepackage[inline]{enumitem}
\pgfplotsset{compat=newest}
\tikzset{
    ultra thick/.style={line width=3.2pt}
}
\setcellgapes{1pt}
\def\BibTeX{{\rm B\kern-.05em{\sc i\kern-.025em b}\kern-.08em
    T\kern-.1667em\lower.7ex\hbox{E}\kern-.125emX}}
\newcommand{\CPU}{\mathtt{CPU}}
\newcommand{\MEM}{\mathtt{MEM}}
\newcommand{\STOR}{\mathtt{STOR}}
\newcommand{\LAT}{\mathtt{LAT}}
\newcommand{\BW}{\mathtt{BW}}
\newcommand{\SOURCE}{\mathtt{src}}
\newcommand{\SINK}{\mathtt{snk}}
\newcommand{\STREAM}{\mathit{data}}
\newcommand{\ELEM}{\STREAM}
\newcommand{\SIZE}{\mathtt{Size}}
\newcommand{\Resd}{\mathtt{ResdBW}}
\newcommand{\ALLOC}{\mathtt{alloc}}
\newcommand{\RPL}{\mathtt{RPL}}
\newcommand{\MPL}{\mathtt{MPL}}

\begin{document}
\title{A Two-Sided Matching Model for Data Stream Processing in the Cloud -- Fog Continuum}

\author{\IEEEauthorblockN{ 
{Narges Mehran},
{Dragi Kimovski},
{Radu Prodan}}
\IEEEauthorblockN{Institute of Information Technology, Alpen-Adria-Universit{\"a}t Klagenfurt, Austria\\
Email: \{name\}.\{surname\}@aau.at}
}

\maketitle

\begin{abstract}
Latency-sensitive and bandwidth-intensive stream processing applications are dominant traffic generators over the Internet network. A stream consists of a continuous sequence of data elements, which require processing in nearly real-time. To improve communication latency and reduce the network congestion, Fog computing complements the Cloud services by moving the computation towards the edge of the network. Unfortunately, the heterogeneity of the new Cloud -- Fog continuum raises important challenges related to deploying and executing data stream applications. We explore in this work a two-sided stable matching model called Cloud -- Fog to data stream application matching (CODA) for deploying a distributed application represented as a workflow of stream processing microservices on heterogeneous computing continuum resources. In CODA, the application microservices rank the continuum resources based on their microservice stream processing time, while resources rank the stream processing microservices based on their residual bandwidth. A stable many-to-one matching algorithm assigns microservices to resources based on their mutual preferences, aiming to optimize the complete stream processing time on the application side, and the total streaming traffic on the resource side.
We evaluate the CODA algorithm using simulated and real-world Cloud -- Fog experimental scenarios. We achieved \SIrange{11}{45}{}\% lower stream processing time and \SIrange{1.3}{20}{}\% lower streaming traffic compared to related state-of-the-art approaches. 

\end{abstract}
\textcolor{red}{\scriptsize 2021 IEEE.  Personal use of this material is permitted.  Permission from IEEE must be obtained for all other uses, in any current or future media, including reprinting/republishing this material for advertising or promotional purposes, creating new collective works, for resale or redistribution to servers or lists, or reuse of any copyrighted component of this work in other works.}

\section{Introduction} \label{sec:intro}
The world is witnessing an exponential growth in the amount of generated data in the presence of pervasive Internet connectivity. Latency-sensitive and bandwidth-intensive data stream processing services, such as live video and video-on-demand streams, are amongst the dominating high velocity traffic generators in today's world. Processing such data streams in nearly real-time~\cite{lai2018optimal} requires vast amounts of computational and network resources in proximity of the data sources. 
However, the high communication penalty for reaching the Cloud data centers significantly hinders the timely processing of the data streams~\cite{mortazavi2017cloudpath, aral2019addressing}.
Fog computing complements the Cloud services by moving the computation towards the edge of network. The extension of the Cloud with distributed micro-data centers (also called cloudlets~\cite{marin2017we}) and mobile Edge servers~\cite{kekki2018mec} forms the so-called \emph{Cloud -- Fog continuum}, which aids the application execution by improving the communication latency and reducing the network congestion.

However, the heterogeneity of the Cloud -- Fog continuum raises multiple challenges for executing data stream processing applications~\cite{de2018distributed}, including application deployment and resources allocation. Unfortunately, existing works often omit to consider data stream applications with strict latency and bandwidth requirements. It becomes therefore essential to explore models for allocating resources to data stream processing applications in the Cloud -- Fog continuum.

We propose a two-sided matching model called \emph{Cloud -- fOg to Data stream application mAtching (CODA)} to address the problem of deploying data stream processing applications organized as directed acyclic graphs on heterogeneous computing continuum resources. CODA approaches this problem using matching theory principles 
involving two sets of players:
\begin{itemize}[leftmargin=*,align=left]
    \item \emph{Application microservices} rank the continuum resources based on their microservice stream processing time (also referred in the following as microservice time);
    \item \emph{Cloud -- Fog resources} rank the stream processing microservices based on their residual bandwidth. 
\end{itemize}
The CODA two-sided stable matching model assigns microservices to resources based on their mutual preferences, aiming to optimize the stream processing time on the application side, and the total streaming traffic on the resource side~\cite{abedin2018resource}.

Hence, the main contributions in this work are:
\begin{itemize}[leftmargin=*,align=left]
    \item A model for quantifying the microservice stream processing time and the residual network bandwidth to a resource;
    \item A ranking strategy tailored to data stream applications that avoids zero bandwidth surplus;
    \item A many-to-one matching model that allocates resources based on their capacity to multiple microservices;
    \item A two-sided stable matching model for allocating Cloud -- Fog resources to a microservice-based data stream processing application.
\end{itemize}

The paper has eight sections. Section \ref{sec:related} surveys the relevant related work. Section \ref{sec:model} elaborates the model underneath our approach, followed by the CODA matching algorithm in Section \ref{sec:alg}. Section \ref{sec:casestudy} describes the stream processing case study application, evaluated using simulation in Section \ref{sec:sim}. Section \ref{sec:testbed} confirms the simulation in a real-world testbed and Section \ref{sec:concl} concludes the paper.

\section{Related work} \label{sec:related}
This section reviews the state-of-the-art in Cloud -- Fog resource allocation for data stream processing applications with reduced network streaming traffic.

\paragraph{Hierarchical resource allocation} Gupta et al.~\cite{gupta2017ifogsim} proposed a hierarchical placement strategy
that executes the last microservice of every application in the Cloud and places all its predecessors on the less powerful computational resources in the Cloud -- Fog hierarchy. Similarly, Mortazavi et al.~\cite{mortazavi2017cloudpath} presented a novel paradigm called CloudPath computing that enables data stream processing on a progression of Cloud data centers based on their computing and storage capabilities, interposed along the geographical span of the network.

\paragraph{Stream processing time reduction} Sharghivand et al.~\cite{sharghivand2018qos} proposed a two-sided matching model for allocating Fog resources to services at the edge of network considering the service response time. The approach improves the user satisfaction and quality of experience using a set of heterogeneous quality of service metrics. Cai et al.~\cite{cai2018response} also addressed the service response time by defining a placement optimization model for complex event-processing applications on Edge resources. The proposed approximation algorithm deploys the operators on the Edge infrastructure with the lowest predicted delay. Veith et al.~\cite{da2018latency} proposed a placement strategy called \mbox{RTR-RP}, which uses a greedy strategy to identify the resources that minimize the service response time by reducing the end-to-end event latency of a data stream analytic application. This approach decomposes the application in data processing flow patterns such as fork and join, and then distributes it on the Cloud -- Fog continuum. Dautov et al.~\cite{dautov2020stream} describes a new approach for stream data processing in Fog by supporting run-time clustering of heterogeneous low powered devices. Besides, they utilise horizontal offloading of computational tasks between the Fog devices, which results in reduction of the communication latency by a factor of five compared to the vertical offloading approaches that rely on the Cloud.

\paragraph{Streaming traffic reduction}
Aral et al.~\cite{aral2019addressing} considered the Fog computing characteristics
to improve the user experience for latency-sensitive applications. The service placement evaluates the network quality of each Cloud and Fog node with respect to its requirements, in particular the connectivity and bandwidth.
Zamani et al.~\cite{zamani2017deadline} describe a semi-real time data stream processing approach at the edge of the network, which supports stream transformation and analysis from source to destination. The approach leverages an in-network computational model that employs software defined networks to dynamically establish data stream routes that exploit the underutilized computational resources at the Edge.

\paragraph{CODA contribution} These works investigate the resource allocation as an optimization problem that minimizes the stream processing time as a main objective, and neglect the streaming traffic
. We extend the related approaches by researching a novel resource provisioning approach based on two-sided matching~\cite{ehsanpour2019efficient} that considers different interests of the involved stakeholders:
\begin{enumerate}
\item minimization of the stream processing time from the application perspective;
\item minimization of traffic considering changes in the data stream rate from the resource provider perspective.
\end{enumerate}

\section{Model} \label{sec:model}
This section presents a formal model and a set of essential definitions important for this work.

\subsection{Stream application}  \label{ssec:AppModel}
We model a \emph{data stream processing application}: \[\mathcal{A} = \left(\mathcal{M}, \mathcal{E}, m_{\SOURCE}, m_{\SINK}, \SOURCE, \SINK \right)\] as a directed acyclic graph (DAG) consisting of:
\begin{enumerate}[align=left, leftmargin=*]
\item A set of $\mathcal{N}_{\mathcal{M}}$ lightweight interconnected \emph{microservices}: \[\mathcal{M} =\left\{m_i\ |\ 0 \leq i < \mathcal{N}_{\mathcal{M}}\right\};\]
\item A set of \emph{data streams} $\STREAM_{ui}$ flowing from an upstream microservice $m_u$ to a downstream microservice  $m_i \in \mathcal{M}$: \[\mathcal{E} = \left\{ \left(m_u, m_i, \STREAM_{ui}\right) | \left(m_u, m_i\right) \in \mathcal{M} \times \mathcal{M}\right\};\]
\item A \emph{source microservice} $m_{\SOURCE}$ processing the data stream produced by $\SOURCE$ of the application $\mathcal{A}$. The source microservice has no upstream microservices:
\[\left(m_{\SOURCE}, m_i, \SOURCE\right) \in \mathcal{E}\\ \land \not\exists \left( m_i, m_{\SOURCE}, \_\right) \in \mathcal{E};\]
\item A \emph{sink microservice} $m_{\SINK}$ generating the data stream for $\SINK$, representing the output of the application $\mathcal{A}$.
\[\left(m_i, m_{\SINK}, \SINK\right) \in \mathcal{E}\\ \land \not\exists \left(m_{\SINK}, m_i, \_\right) \in \mathcal{E}.\]
\end{enumerate}

We define a \emph{data stream} using the following triple notation: $\STREAM_{ui} = \left(\ELEM_{ui}[x], \lambda_{ui}, \SIZE_{ui}\right)$, where:
\begin{enumerate}[align=left, leftmargin=*]
\item $\ELEM_{ui}$ represents a sequence of data stream elements sent between two microservices $m_u$ and $m_i$, measured in \SI{}{\bit};
\item $\ELEM_{ui}[x]$ is a single data element in the data stream $\ELEM_{ui}$. We assume that $m_i$ recognizes the data elements in the stream $\ELEM_{ui}$ by the timestamp and merges the elements in the correct order;
\item $\lambda_{ui}$ represents the ingress data rate that the microservice $m_i$ receives a number of data elements per unit of time from its upstream microservice $m_u$~\cite{zhao2016pricing}.
\item $\SIZE_{ui}$ is the total number of data elements $\ELEM_{ui}[x]$ transmitted in a stream $\STREAM_{ui}$, where $1 \leq x \leq \SIZE_{ui}$.
\end{enumerate}

Proper processing of a data element $\ELEM_{ui}[x]$ by a microservice $m_{i}$ has certain \emph{resource requirements} in terms of the processing load $\CPU\left(m_i, \ELEM_{ui}[x]\right)$ (measured in million of instructions (MI)), memory $\MEM\left(m_i,\ELEM_{ui}[x]\right)$ and storage $\STOR\left(m_i,\ELEM_{ui}[x]\right)$ (measured in \SI{}{\mega\byte}).

\subsection{Resource model}  \label{ssec:ResModel}
We define a Cloud -- Fog environment as a set of $\mathcal{N}_{\mathcal{R}}$ \emph{resources}: $\mathcal{R}=\left\{r_j\ |\ 0 \leq j < \mathcal{N}_{\mathcal{R}}\right\}$, where a resource $r_j$ defines its computational power $\CPU_j$ (in MI per second), memory size $\MEM_j$, and storage size $\STOR_j$: $r_{j} = \left(\CPU_j, \MEM_j, \STOR_j, c_j\right).$

We define the \emph{capacity} $c_j$ of a resource $r_j$ as the maximum number of microservices it can host, which relies on its utilization as a threshold~\cite{mao2017draps,birkenheuer2008virtual} 
that ensures no contention among the microservices~\cite{VirtualizationViaContainers}.

We  model the \emph{network channels} between the Cloud -- Fog resources as $\mathcal{L}=\{l_{qj}\ |\ 0 \leq q,j < \mathcal{N}_{\mathcal{R}}\}$, where $l_{qj}=\left(\LAT_{qj}, \BW_{qj} \right)$ represents by the round-trip latency $\LAT_{qj}$ and network bandwidth $\BW_{qj}$ between the resources $r_q$ and $r_j$. Two interdependent microservices allocated to the same resource have $\LAT_{qj} = 0$ and $\BW_{qj} = \infty$~\cite{DEMAIO2020171}.

We define a \emph{microservice allocation} as a mapping function $\mu: \mathcal{A} \to \mathcal{R}$ that assigns a microservice $m_i$ to a resource $r_j = \mu\left(m_i\right)$.
Accordingly, $\ALLOC(r_j)$ represents the list of \emph{microservices} allocated and deployed on each resource $r_j$:
\[\ALLOC(r_j) = \left\{ m_i\ |\ \mu(m_i) = r_j\right\}.\]

\subsection{Ranking methods} \label{ssec:rank}
The CODA model for matching application microservices to resources uses a two-sided ranking method:
\begin{itemize}[align=left, leftmargin=*]
    \item microservice-side ranking that considers the stream processing time of each microservice;
    \item resource-side ranking that considers the residual bandwidth to each resource allocated to each microservice.
\end{itemize}

\subsubsection{Microservice-side ranking} \label{sssec:msrank}
We define the \emph{element processing time} $t\left(m_i, \ELEM_{ui}[x], r_j\right)$ required by a microservice $m_i$ to process the $x^{\mathit{th}}$ element $\ELEM_{ui}[x]$ of a stream received by a resource $r_j\ = \mu\left(m_i\right)$ as the sum of three terms:
\begin{multline*}
t\left(m_i, \ELEM_{ui}[x], r_j\right)= \frac{\CPU\left(m_i,\ELEM_{ui}[x]\right)}{\CPU_j}+\\
+\frac{\ELEM_{ui}[x]}{\BW_{qj}}+\LAT_{qj},
\label{eq:mstimeforelem}
\end{multline*}
\paragraph{computation time} as the ratio between the computational requirement $CPU\left(m_i,\ELEM_{ui}[x]\right)$ for processing a data element $\ELEM_{ui}[x]$ on the microservice $m_i$ and the processing speed $\CPU_j$ of the resource $r_j$;
\paragraph{transmission time} as the ratio between the size of the received data element $\ELEM_{ui}[x]$ and the network bandwidth $\BW_{qj}$ to $r_j = \mu\left(m_i\right)$~\cite{mehran2019mapo};
\paragraph{latency} as the round-trip time $\LAT_{qj}$ between resources $r_q$ and $r_j$.

The \emph{microservice stream processing time} $T\left(m_i, \STREAM_{ui}, r_j\right)$ of a data stream $\STREAM_{ui}$ processed by a microservice $m_i$ running on a resource $r_j$ is the sum of its element processing times $t\left(m_i, \ELEM_{ui}[x], r_j\right)$:
\begin{equation*}
    T\left(m_i, \STREAM_{ui}, r_j\right) =\\
    \sum\limits_{x=1}^{\SIZE_{ui}}
    t\left(m_i, \ELEM_{ui}[x], r_j\right).
    \label{eq:mstimeforstream}
\end{equation*}
Every microservice $m_i$ ranks the resources in a \emph{resource preference list} $\RPL[m_i]$ based on the microservice stream processing time,
as presented in Algorithm~\ref{alg:msrank}. The resource that guarantees a lower microservice time receives a higher rank.
The algorithm first initializes the resource preference lists for each microservice with the empty set in line~\ref{alg:msinit}.
Thereafter, it filters the resources that do not satisfy the memory $\MEM\left(m_i,\ELEM_{ui}[x]\right)$ and storage $\STOR\left(m_i,\ELEM_{ui}[x]\right)$ requirements of a microservice (line~\ref{alg:checkmemconstraint}). Afterward, it creates a list of tuples for each microservice $m_i$ that associates the maximum microservice time $T\left(m_i, \STREAM_{ui}, r_j\right)$ of all upstream microservices $m_u$ of $m_i$ to each resource $r_j$ (line~\ref{alg:mstimeforstream}).
Finally, the algorithm sorts the resource preferences of each microservice based on its microservice time in descending order in line~\ref{alg:ms-side-ranking}.

\begin{algorithm}[t]
\caption{Microservice-side ranking algorithm.}
\label{alg:msrank}
\scriptsize
\textbf{Input:} $\mathcal{A} = \left(\mathcal{M}, \mathcal{E}, m_{\SOURCE}, m_{\SINK}, \SOURCE, \SINK \right)$\Comment{Stream application}\\
\hspace*{\algorithmicindent}\hspace*{\algorithmicindent} $\mathcal{R}=\{r_j\ |\ 0 \leq j < \mathcal{N}_{\mathcal{R}}\}$\Comment{Cloud -- Fog resource set}\\
\hspace*{\algorithmicindent}\hspace*{\algorithmicindent} 
$\mathcal{L}=\{l_{qj}\ |\ 0 \leq q,j < \mathcal{N}_{\mathcal{R}}\}$\Comment{Cloud -- Fog channel set}\\
\textbf{Output:} $\RPL[m_i], \forall m_i \in \mathcal{M}$\Comment{Resource preference lists of all microservices $m_i$}
\begin{algorithmic}[1]
\ForAll{$m_i\in \mathcal{M}$}\label{alg:msinit}\Comment{Initialize $\RPL$}
    \State $\RPL[m_i] \gets \emptyset$
\EndFor
\ForAll{$m_i \in \mathcal{M}$}
    \ForAll{$\left(r_j \in \mathcal{R}\right)\ \land\ \left(l_{qj} \in \mathcal{L}\right)$}
        \If{$\left(\MEM\left(m_i\right) < \MEM_j\right)\land \left(\STOR\left(m_i\right) < \STOR_j\right)$}\Comment{Check constraints}\label{alg:checkmemconstraint}
            \State{$\mathit{\RPL}[m_i] \gets \RPL[m_i] \bigcup (r_j, \max\limits_{\substack{\forall (m_u, m_i,\\ \STREAM_{ui}) \in \mathcal{E}}} T(m_i, \STREAM_{ui}, r_j))$}\label{alg:mstimeforstream}
        \EndIf\Comment{Add $r_j$ and its microservice time to $m_i$'s $\RPL$}
    \EndFor
\EndFor
\ForAll {$\left(m_i \in \mathcal{M}\right)\ \land\ \left(\mathit{\RPL}[m_i]\neq \emptyset\right)$}
    \State{$\mathit{\RPL}[m_i] \gets Sort_{T}(\mathit{\RPL}[m_i])$}\Comment{Sort tuples based on microservice time}\label{alg:ms-side-ranking}
\EndFor
\State \Return $\RPL$;
\end{algorithmic}
\end{algorithm}

\subsubsection{Resource-side ranking} \label{sssec:resrank}
We model the \emph{residual bandwidth} to a resource $r_j$ as 
the difference between the available bandwidth $\BW_{qj}$ and the ingress traffic from an upstream microservice, as defined in the DAG structure of the applications. The ingress traffic is the amount of data per time unit received by a resource $r_j$ allocated to a microservice $m_i$, which depends on ingress data rate $\lambda_{ui}$ and data stream $\ELEM_{ui}$:
\begin{equation*}
\scriptsize
\Resd_{j}\left(m_i, \STREAM_{ui}, r_j\right) = \BW_{qj} -\sum\limits_{x=1}^{\SIZE_{ui}} \left(\lambda_{ui}\cdot  \ELEM_{ui}[x]\right).
\label{eq:resrank}
\end{equation*}
The resource-side ranking, presented in Algorithm~\ref{alg:resrank}, receives as input the resource preference lists $\mathit{\RPL[m_i]}\ (\forall m_i \in \mathcal{A})$ computed in Algorithm~\ref{alg:msrank}, along with the application $\mathcal{A}$, the resource set $\mathcal{R}$, and the set of network channels $\mathcal{L}$. Similarly, the algorithm initializes the \emph{microservice preference list} $\MPL[r_j]$ of each resource with the empty set in line~\ref{alg:resinit}.
Afterward, each resource ranks the microservices in a preference list in line~\ref{alg:resdbw} based on its residual bandwidth.
Finally, the algorithm sorts the microservice preferences in descending order in line~\ref{alg:res-side-ranking} based on the residual bandwidth.
Hence, the microservice that offers a lower bandwidth utilization receives a higher rank.

\begin{algorithm}[t]
\caption{Resource-side ranking algorithm.}
\label{alg:resrank}
\scriptsize
\textbf{Input:} $\mathcal{A} = \left(\mathcal{M}, \mathcal{E}, m_{\SOURCE}, m_{\SINK}, \SOURCE, \SINK \right)$,\Comment{Stream app.}\\
\hspace*{\algorithmicindent}\hspace*{\algorithmicindent}
$\mathcal{R}=\{r_j\ |\ 0 \leq j < \mathcal{N}_{\mathcal{R}}\}$\Comment{Cloud -- Fog resource set}\\
\hspace*{\algorithmicindent}\hspace*{\algorithmicindent} 
$\RPL[m_i], \forall m_i \in \mathcal{M}$\Comment{Resource preference lists of all microservices $m_i$}\\
\hspace*{\algorithmicindent}\hspace*{\algorithmicindent} 
$\mathcal{L}=\{l_{qj}\ |\ 0 \leq q,j < \mathcal{N}_{\mathcal{R}}\}$\Comment{Cloud -- Fog channel set}\\
\textbf{Output:} $\MPL[r_j]$, $\forall r_j \in \mathcal{R}$\Comment{Microservice preference lists of all resources $r_j$}
\begin{algorithmic}[1]
\ForAll{$r_j\in \mathcal{R}$}\label{alg:resinit}\Comment{Initialize $\MPL$}
    \State $\MPL[r_j] \gets \emptyset$
\EndFor
\ForAll{$m_i \in \mathcal{M}$}
    \ForAll{$ \left(r_j  \in \RPL[m_i]\right)\ \land\ \left(l_{qj} \in \mathcal{L}\right)$}
            \State{$\MPL[r_j] \gets \MPL[r_j]\bigcup \left(m_i, \Resd_{j}\right)$}\label{alg:resdbw}
    \EndFor\Comment{Add $m_i$ and its residual bandwidth to $r_j$'s $\mathit{\MPL}$}
\EndFor
\ForAll{$\left(r_j \in \mathcal{R}\right)\ \land\ \left(\mathit{\MPL}[r_j]\neq \emptyset\right)$}
    \State{$\MPL[r_j] \gets Sort_{\Resd}(\MPL[r_j])$}
    \Comment{Sort tuples based on residual bandwidth}\label{alg:res-side-ranking}
\EndFor
\State \Return $\MPL$;
\end{algorithmic}
\end{algorithm}

\subsection{Problem definition} \label{ssec:probdef}
Matching theory is a formal framework describing the interactions among interdependent rational entities and forming mutually beneficial relationships over time~\cite{bayat2016matching}. The analytical matching theory helps to assign a set of rational entities to one another, typically subject to constraints such as preference lists and capacities~\cite{gale1962college}. 

We represent our \emph{resource allocation problem} as a matching game using two finite and disjoint sets of players:
\begin{enumerate*}
\item the microservices $\mathcal{M}$ of the stream processing application $\mathcal{A}$, and
    \item the Cloud -- Fog resources in $\mathcal{R}$.
\end{enumerate*}
The game aims to match each microservice $m_i \in \mathcal{M}$ to a resource in $r_j \in \mathcal{R}$ with sufficient capacity that optimizes \emph{two independent goals}:
\begin{enumerate*}
\item application-specific on one side and
\item resource provider-specific on the other side. \end{enumerate*}
The result is a bilateral resource allocation agreement that represents the players' preferences over each other.
Section~\ref{subsec:metrics} instantiates this problem on two metrics:
\begin{enumerate*}
\item \emph{stream processing time} on the application side, and
\item \emph{total streaming traffic} on the resource side.
\end{enumerate*}

In a matching game, a microservice $m_i \in \mathcal{M}$ asks for allocation on the first resource $r_j$ in its preference list $\RPL[m_i]$. If $r_j$ has enough capacity $c_j$ and there exists no other preferred microservice in its preference list $\MPL[r_j]$, it bids for $m_i$. If the two sides agree, the microservice $m_i$ holds its demand from the resource $r_j$ and vice versa until the matching completes.

A \emph{valid} resource allocation is (pairwise) stable if it satisfies three properties of a many-to-one matching game~\cite{diebold2014course,wilde2020novel}:
\begin{enumerate}[align=left, leftmargin=*]
    \item Each microservice is allocated to exactly one resource from its preference list:
    \[\mu(m_i)\in\mathcal{R}\ \land\ \left|\mu\left(m_i\right)\right| = 1\ \land\ \mu(m_i)\in \RPL[m_i];\]
    \item A resource can host multiple microservices that are part of its preference list and within its capacity:
    \[\ALLOC\left(r_j\right) \subseteq \MPL\left[r_j\right]\subseteq\mathcal{M}\ \land\ \left|\ALLOC\left(r_j\right)\right| \leqslant c_j;\]
    \item The matching does not contain blocking pairs of microservices and resources that prefer matching each other rather than their current assignments~\cite{abedin2018resource}.
    A matching $r_j = \mu\left(m_i\right)$ is \emph{not blocking} if the following conditions hold:
    \begin{enumerate}
        \item $m_i$ and $r_j$ are currently matched with each other;
        \item $m_i$ does not prefer another resource to its current matching $r_j$;
        \item $r_j$ does not prefer another microservice to any of its current matching in $\ALLOC(r_j)$.
    \end{enumerate}
\end{enumerate}


\section{CODA matching algorithm}~\label{sec:alg}
Algorithm~\ref{alg:matchingMany2One} describes the many-to-one matching-based allocation of microservices to resources.
The algorithm receives as input the stream application described as a DAG, the resource preference list $\RPL$ of each microservice $m_i$, and the microservice preference list $\MPL$ of each resource $r_j$, computed by Algorithms~\ref{alg:msrank} and~\ref{alg:resrank}. The algorithm outputs a stable matching between microservices and resources $\mu(\mathcal{A}) \subseteq \mathcal{R}$. After initializing the allocation on both sides (lines~\ref{startinit}--\ref{endinit}), the algorithm loops until it manages to find the appropriate resource allocation matches to all microservices according to their mutual preferences (lines~\ref{startwhile}--\ref{endwhile}). In every iteration, it attempts to find a good resource matching for every microservice using several matching states (i.e. State-1, State-2.1, State-2.2), described in the following paragraphs.

\subsubsection{State-1}\label{sss:state1} Each microservice not yet matched to any resource demands the resource $r_j$ with the lowest microservice time, ranked first in its preference list $\RPL[m_i]$ (lines~\ref{startwhile}--\ref{TakeAnyUnmatchedResource}). If the resource $r_j$ has also ranked the microservice first in its preference list $\MPL\left[r_j\right]$ because of the least bandwidth consumption (line~\ref{rFirst?}), the algorithm creates a matching pair and update the resource $\mu\left(m_i\right)$ and the list of microservices $\ALLOC\left(r_j\right)$ (lines~\ref{mFirstR}--\ref{mFirstM}).

\subsubsection{State-2}\label{sss:state2} If the microservice $m_i$ is not the first in the preference list $\MPL\left[r_j\right]$ (line~\ref{rankedS?}), $m_i$ matches to $r_j$ (lines~\ref{MatchMs}--\ref{MatchRe}). Afterward, the algorithm checks the following two states:

\paragraph{State-2.1}\label{sss:state21} If $m_i$'s allocation to resource $r_j$ exceeds its capacity $c_j$ (line~\ref{overC}), the algorithm removes the allocation $\mu\left(m_u\right)=r_j$ with the lowest residual bandwidth in the ranked preference list $\MPL\left[r_j\right]$ of resource $r_j$ (lines~\ref{unMs}--\ref{unMe}).

\paragraph{State-2.2}\label{sss:state22} If a resource $r_j$ reaches its capacity $c_j$ (line~\ref{capF}), the algorithm identifies the microservice $m_u$ with the lowest residual bandwidth in its allocation list $\ALLOC\left(r_j\right)$ (line~\ref{last}). Afterward, $r_j$ removes all microservices $m_s$ with a lower residual bandwidth than $m_u$ from its preference list $\MPL\left[r_j\right]$. Similarly, all microservices $m_s$ remove $r_j$ from their resource preference lists $\RPL[m_s]$. This avoids deploying microservices with low residual bandwidth on $r_j$ and allows higher ranked microservices in $\MPL\left[r_j\right]$ to fill its capacity (lines~\ref{higherrank}--\ref{delfrommpl}).

\begin{algorithm}[t]
\scriptsize
\caption{CODA matching algorithm.}
\label{alg:matchingMany2One}
\textbf{Input:} $\mathcal{A} = \left(\mathcal{M}, \mathcal{E}, m_{\SOURCE}, m_{\SINK}, \SOURCE, \SINK \right)$,\Comment{Stream application}\\
\hspace*{\algorithmicindent}\hspace*{\algorithmicindent}
$\mathcal{R}=\left\{r_j\ |\ 0 \leq j < \mathcal{N}_{\mathcal{R}}\right\}$\Comment{Cloud -- Fog resource set}\\
\hspace*{\algorithmicindent}\hspace*{\algorithmicindent} 
$\RPL\left[m_i\right], \forall m_i \in \mathcal{M}$\Comment{Preference lists of all microservices $m_i$}\\
\hspace*{\algorithmicindent}\hspace*{\algorithmicindent} 
$\MPL\left[r_j\right], \forall r_j \in \mathcal{R}$\Comment{Preference lists of all resources $r_j$}\\
\textbf{Output:} $\mathcal{R} = \mu(\mathcal{A})$.
\begin{algorithmic}[1]
\ForAll{$m_i\in \mathcal{M}$}\Comment{Initialize invalid microservice allocation}\label{startinit}
    \State{$\mu\left(m_i\right)\gets \mathit{NaR}$}\Comment{Not a Resource}
\EndFor
\ForAll{$r_j \in \mathcal{R}$} \Comment{Initialize empty resource allocation}
    \State{$\ALLOC\left(r_j\right)\gets \emptyset$}
\EndFor\label{endinit}
\State{$NaR\_\mathcal{M} \gets \mathcal{M}$}\Comment{Initialize list of Not-a-Resource microservices}
\While{$NaR\_\mathcal{M} \neq \emptyset$}\Comment{Allocate all microservices}\label{startwhile}
            \State{$m_i\gets \Call{First}{NaR\_\mathcal{M}}$}\Comment{ list of Not-a-Resource}
            \State{$r_j\gets \Call{First}{\RPL\left[m_i\right]}$}\label{TakeAnyUnmatchedResource}
            \If{$m_i =  \Call{First}{\MPL\left[r_j\right]} \land \left|\ALLOC\left(r_j\right)\right| \neq c_j$} \Comment{\emph{State-1}}\label{rFirst?}
                \State{$\mu\left(m_i\right)\ \gets\  r_j$}\Comment{Match $m_i$ and $r_j$}\label{mFirstR}
                \State{$\ALLOC\left(r_j\right)\ \gets\  \Call{Sort$_\Resd$}{\ALLOC\left(r_j\right) \cup m_i}$}\label{mFirstM}
            \Else
            \If{$m_i\ \in \ \MPL\left[r_j\right] \land m_i \neq \Call{First}{ \MPL\left[r_j\right]}$}\label{rankedS?}\Comment{\emph{State-2}}
                \State{$\mu\left(m_i\right)\ \gets\  r_j$}\Comment{Match $m_i$ and $r_j$}\label{MatchMs}
                \State{$\ALLOC\left(r_j\right)\ \gets\  \Call{Sort$_\Resd$}{\ALLOC\left(r_j\right) \cup m_i}$}\label{MatchRe}
                \If{$|\ALLOC\left(r_j\right)|\ >\ c_j$}\Comment{\emph{State-2.1}}
                \label{overC}
                    \State{$m_u \gets \Call{Last}{\ALLOC\left(r_j\right)}$}\label{unMs}
                    \State{$\mu\left(m_u\right)\ \gets\  \mathit{NaR}$}\Comment{Reject $m_i$ from $r_j$}
                    \State{$\ALLOC\left(r_j\right)\ \gets\  \ALLOC(r_j)\setminus {m_u}$}\label{unMe}
                    \State{$NaR\_\mathcal{M}.\Call{Append}{m_u}$}\Comment{Add $m_u$ to list of Not-a-Resource}
                \EndIf
                \If{ $\left|\ALLOC\left(r_j\right)\right| = c_j$}\Comment{\emph{State-2.2}}\label{capF}
                    \State{$m_u \gets \Call{Last}{\ALLOC\left(r_j\right)}$} \label{last}
                    \ForAll{$m_s\in \MPL[r_j] \land \Call{Rank}{m_u} < \Call{Rank}{m_s}$}\label{higherrank}
                        \State{$\RPL\left[m_s\right]\ \gets \ \RPL[m_s] \setminus {r_j}$}\label{delfromrpl}
                        \State{$\MPL\left[r_j\right]\ \gets \ \MPL\left[r_j\right] \setminus {m_s}$}\label{delfrommpl}
                        \If{$\RPL[m_s] = \emptyset$}
                            \State{$NaR\_\mathcal{M}.\Call{Remove}{m_s}$}\Comment{Remove $m_s$ if it has no prefs}
                        \EndIf
                    \EndFor
                \EndIf
            \EndIf
        \EndIf
\EndWhile\label{endwhile}
\end{algorithmic}
\end{algorithm}

\subsubsection{CODA complexity}
The microservice-side ranking (Algorithm~\ref{alg:msrank}) and the resource-side ranking (Algorithm~\ref{alg:resrank}) algorithms have complexity of $\mathcal{O}\left(\mathcal{N}_{\mathcal{M}}\cdot\mathcal{N}_{\mathcal{R}}\right)$, where $\mathcal{N}_{\mathcal{M}}$ is the number of microservices and $\mathcal{N}_{\mathcal{R}}$ is the number of resources allocated to the microservices.
Algorithm~\ref{alg:matchingMany2One} traverses the microservice $m_i$'s resource preference list $\RPL\left[m_i\right]$ that previously ranked all the resources (outputs of Algorithms~\ref{alg:msrank} and~\ref{alg:resrank}). 
Therefore, its worst-case time complexity directly depends on the number of acceptable matches, which is $\mathcal{N}_{\mathcal{M}} \cdot \mathcal{N}_{\mathcal{R}}$.
The complexity of the sorting algorithm $Sort_{\Resd}$ must consider the maximum capacity of the resources $c_{\max}=\max\left(c_j\right):\ \forall j \leq \mathcal{N}_{\mathcal{R}}$. Considering the use of a quick-sort algorithm, this leads to a total runtime complexity of Algorithm~\ref{alg:matchingMany2One} of $\mathcal{O}\left( c_{\max}\cdot\log\left(c_{max}\right)\cdot \mathcal{N}_{\mathcal{M}} \cdot \mathcal{N}_{\mathcal{R}}\right)$.

\subsubsection{CODA trace example} 
\begin{figure}[t]
\centering
\begin{subfigure}[t]{\columnwidth}
        \includegraphics[width=\columnwidth]{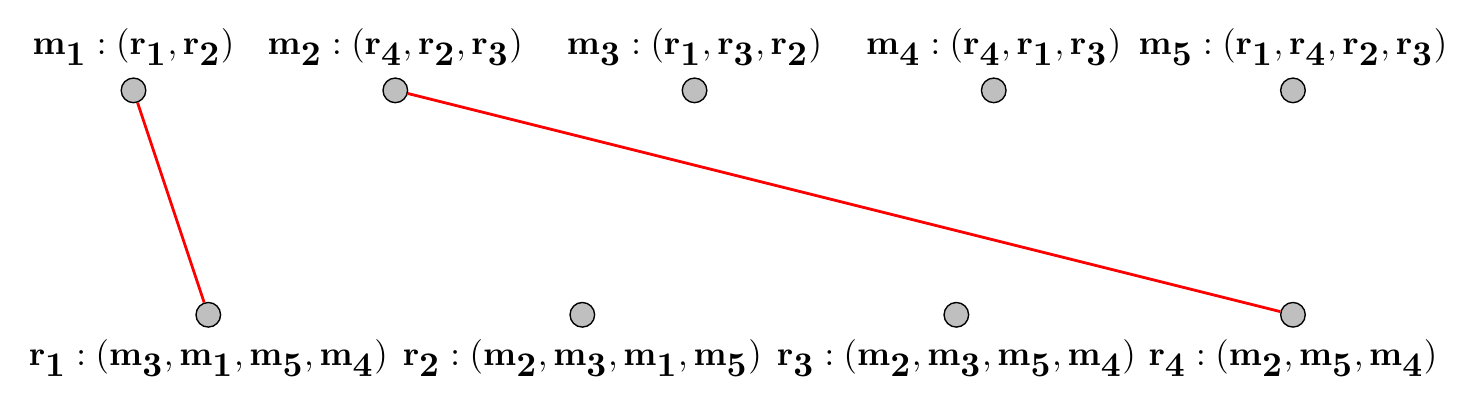}
\caption{State-2: $m_1$ matches to $r_1$; State-1: $m_3$ matches to $r_1$.}
\label{fig:micro2res:a}
\end{subfigure}
\hfill
\begin{subfigure}[t]{\columnwidth}
        \includegraphics[width=\columnwidth]{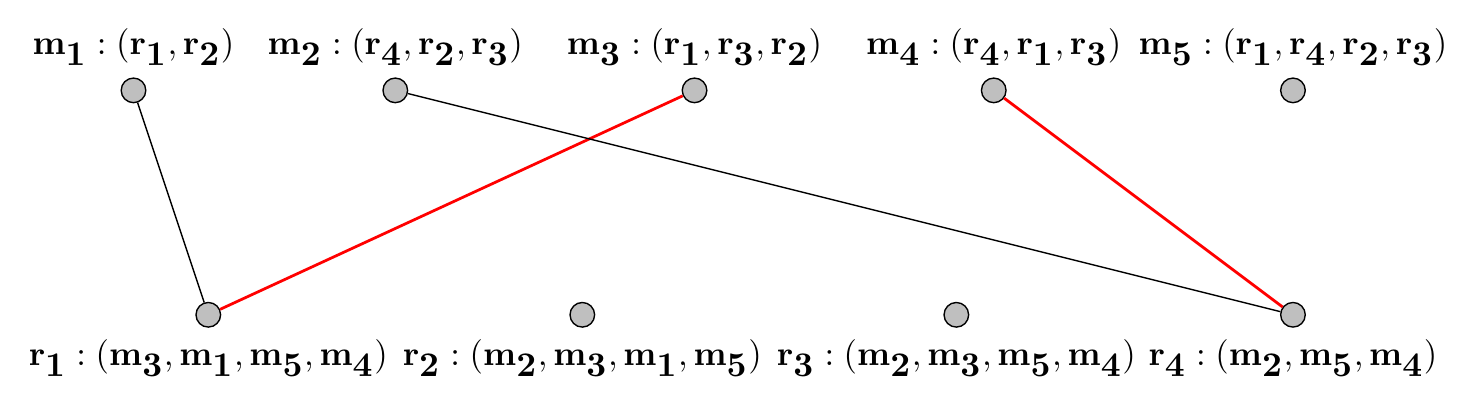}
\caption{State-2: $m_3$ matches to $r_1$; State-2.2: $r_1$ removes $m_4$ and $m_5$; State-2: $m_4$ matches to $r_4$.}
\label{fig:micro2res:b}
\end{subfigure}
\hfill
\begin{subfigure}[t]{\columnwidth}
        \includegraphics[width=\columnwidth]{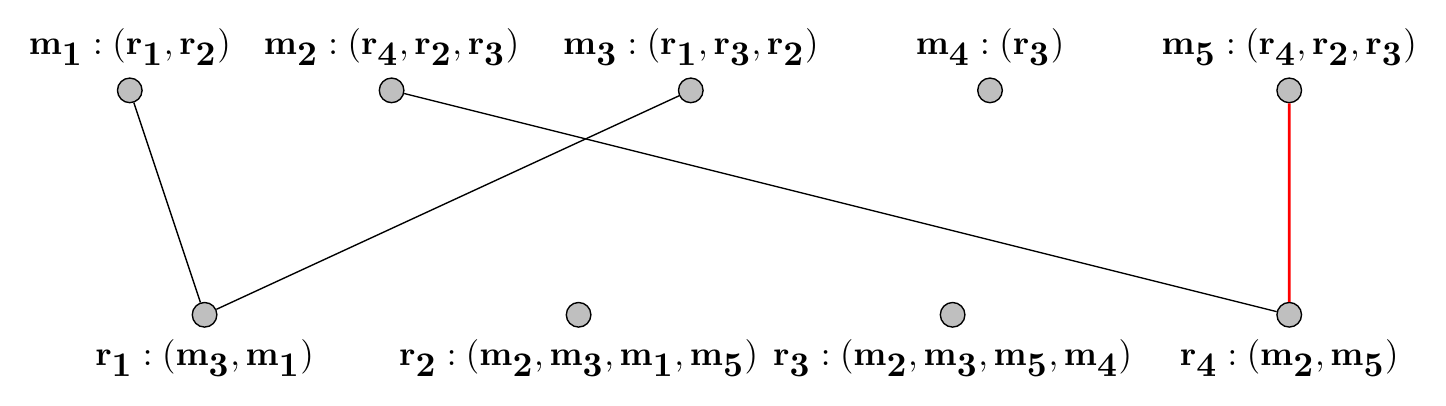}
\caption{State-2: $m_5$ matches to $r_4$; State-2.1: $r_4$ rejects $m_4$; State-2.2: $r_4$ removes $m_4$.}
\label{fig:micro2res:c}
\end{subfigure}
\hfill
\begin{subfigure}[t]{\columnwidth}
        \includegraphics[width=\columnwidth]{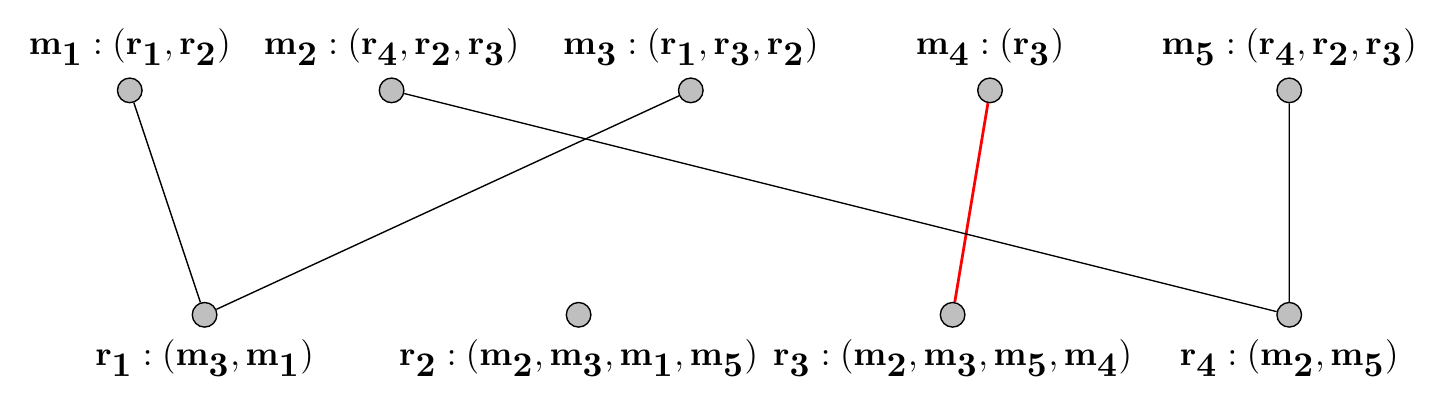}
\caption{State-2: $m_4$ matches to $r_3$.}
\label{fig:micro2res:d}
\end{subfigure}
\caption{Algorithm~\ref{alg:matchingMany2One} trace example (CODA).}
\label{fig:micro2res}
\end{figure}

Figure~\ref{fig:micro2res} illustrates an example of using Algorithm~\ref{alg:matchingMany2One} on five microservices and four resources to converge to a stable matching. Each resource has the capacity to allocate at most two microservices. We assume that Algorithms~\ref{alg:msrank} and~\ref{alg:resrank} already created the microservice and resource preference lists (displayed in brackets in Figure~\ref{fig:micro2res}). Figure~\ref{fig:micro2res:a} shows the matching of the microservice $m_1$ to the resource $r_1$ (lines~\ref{MatchMs}--\ref{MatchRe}). Figure 1a also depicts the matching of $m_2$ to $r_4$ (ranked highest in each other preference lists) following the State-1 of the algorithm (lines~\ref{mFirstR}--\ref{mFirstM}).
Afterward, $m_3$ matches to $r_1$, as shown in Figure~\ref{fig:micro2res:b} (lines~\ref{mFirstR}--\ref{mFirstM}). In addition, Figure~\ref{fig:micro2res:b} illustrates that the microservices with lower residual bandwidth in the resources preference lists (i.e., not ranked first) demand resources, and thus, $m_4$ matches to $r_4$ (lines~\ref{MatchMs}--\ref{MatchRe}). As a consequence, $r_1$ reaches its total capacity and removes the lower-ranked microservices $m_5$ and $m_4$ (with a lower residual bandwidth than $m_1$) from its preference list (State-2.2). The microservices $m_5$ and $m_4$ also remove $r_1$ from their preference lists (lines~\ref{last}--\ref{delfrommpl}). 
In the next outer loop iteration (line~\ref{startwhile}), the remaining microservice $m_5$ demands its preferred resource $r_4$ (line~\ref{TakeAnyUnmatchedResource}). Therefore, $r_4$ and $m_5$ match one another (lines~\ref{MatchMs}--\ref{MatchRe}), although $r_4$ already reached its total capacity. This matching is not successful, as the capacity of $r_4$ is full (State-2.1). However, $m_5$ has a higher rank than $m_4$ due to its higher residual bandwidth; therefore, the matching of $m_4$ to $r_4$ fails (lines~\ref{unMs}--\ref{unMe}). As a consequence, $r_4$ removes $m_4$ (with lower residual bandwidth than $m_5$) from its preference list (State-2.2) and $m_4$ removes $r_4$ from its preference list too (lines~\ref{last}--\ref{delfrommpl}), as shown in Figure~\ref{fig:micro2res:c}.
Finally, Figure~\ref{fig:micro2res:d} shows that the microservice $m_4$ matches resource $r_3$ (lines~\ref{MatchMs}--\ref{MatchRe}), as the only resource with enough capacity that prefers it.

\section{Case study: video stream processing for traffic sign classification} \label{sec:casestudy}
\begin{figure}[!t]
    \centering
    \includegraphics[width=\columnwidth]{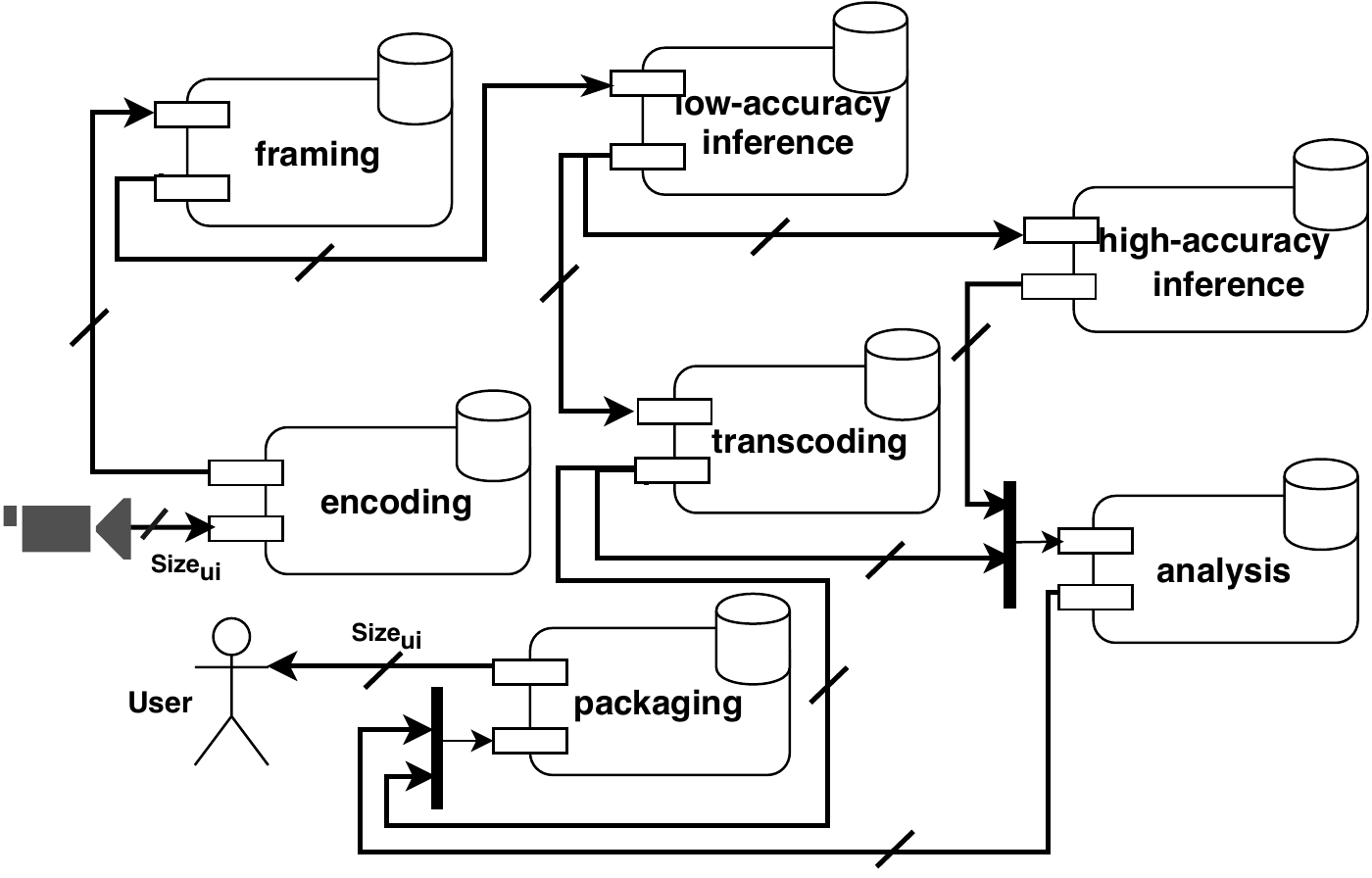}
    \caption{Traffic sign classification in video stream applications.}
    \label{fig:videstreamprocessing-ms}
\end{figure}

We selected a representative traffic management system case study following road safety inspection concerns. Detecting and recognizing different traffic signs and anomalies in nearly real-time requires fast detection of objects in video frames and embedding the information on the detected objects in video streams at different encoding resolutions and bit rates~\cite{vsegvic2014exploiting}. Typical examples are broken, covered, worn-out or stolen traffic signs, or incorrectly painted road surface markings~\cite{vsegvic2010computer}.
We represent this application as a DAG of seven microservices depicted in Figure~\ref{fig:videstreamprocessing-ms}. Each independent microservice contains a data store and communicates with other microservices through a lightweight HTTP interface~\cite{pallewatta2019microservices}. 

\subsubsection{Encoding microservice} receives and encodes the raw video stream in high resolution and bitrate near to the vehicles equipped with multiview-cameras. We use for this purpose the \texttt{ffmpeg} software suite~\cite{zabrovskiy2018multi} with the H.264 video codec for encoding, transcoding and packaging of the video streams.
\subsubsection{Framing microservice} 
utilizes \texttt{OpenCV} to produce still frames from different video scenes~\cite{opencv}.
\subsubsection{Low-accuracy inference microservice} identifies features in the video stream, such as traffic signs on the road. The microservice uses \texttt{TensorFlow} core version $2.3.0$ for \texttt{Python v3.7.4} to train a convolutional neural network with nine layers on localized signs from $50,000$ video frames of $43$ different traffic sign classes. Every frame contains a traffic sign used for training and testing the neural network. This microservice aims for a low classification accuracy of $70\%$.

\subsubsection{High-accuracy inference microservice} uses a machine learning model~\cite{ananthanarayanan2019video} capable of accurate inference when the low-accuracy microservice has a poor confidence. We use the same convolutional neural network with nine layers to classify the signs in the same video frames as for the low-accuracy inference until reaching a $90\%$ accuracy.
\subsubsection{Analysis microservice} updates and retrains the multi-class classification model to learn from newly collected data~\cite{stallkamp2012man}. This microservice is the upstream of the high-accuracy inference and transcoding microservices and requires a barrier to synchronize the received data.
\subsubsection{Transcoding microservice} converts the video in different resolutions and bitrates, and prepares it for delivery. We again use the \texttt{ffmpeg} software suite with the H.264 video codec for transcoding the video streams.
\subsubsection{Packaging and delivery microservice} provides the transcoded video stream together with the detected signs in the format required by the drivers. 
This microservice is the downstream of analysis and transcoding microservices, and uses a barrier to synchronize the data received from its upstream microservices.

We used the Phoronix test suite~\cite{phoronix} to benchmark the application microservices on a set of heterogeneous devices integrated in our testbed, described in Section~\ref{sec:testbed}. Afterward, we identified the requirements of the encoding, transcoding, packaging and inference microservices based on the average device utilization in terms of MI, memory, storage, $\ELEM_{ui}$ (in \SI{}{\mega\byte}) and ingress data rate (in $\SI{}{[\per\second]}$). We summarize the video stream processing application requirements in Table~\ref{tbl:dsa}.

\begin{table}[t]
\centering
\caption{Application resource requirements per microservice.}
\label{tbl:dsa}
\begin{tabular}{|@{ }c@{ }|@{ }c@{ }|@{ }c@{ }|@{ }c@{ }|@{ }c@{ }|@{ }c@{ }|}
\cline{2-6}
\multicolumn{1}{{c|}}{}&$\CPU$ & $\MEM$ & \emph{Storage} & $\ELEM_{ui}$ & $\lambda_{ui}$\\ 
\multicolumn{1}{{c|}}{}&$[MI]\cdot 10^{3}$ & $\SI{}{[\mega\byte]}$ & $\SI{}{[\giga\byte]}$ & $\SI{}{[\mega\byte]}$ & $\SI{}{[\per\second]}$\\ 
\hline
\emph{encoding}&$30-40$ & $300-500$ & $1-5$ & $0.1-10$ & $0.2-40$\\
\hline
\emph{framing}&$1-5$ & $100-300$ & $0.5-2$ & $0.1-10$ & $0.2-40$\\
\hline
\emph{low-accuracy inf.}&$5-20$ & $200-500$ & $0.5-2$ & $0.1-10$ & $0.2-40$\\
\hline
\emph{high-accuracy inf.}&$30-40$ & $300-500$ & $3-5$ & $0.1-10$ & $0.2-40$\\
\hline
\emph{analysis}&$10-20$ & $100-300$ & $1-3$ & $0.1-10$ & $0.2-40$\\
\hline
\emph{transcoding}&$5-40$ & $200-500$ & $0.5-5$ & $0.1-10$ & $0.2-40$\\
\hline
\emph{packaging}&$10-20$ & $100-300$ & $1-2$ & $0.1-10$ & $0.2-40$\\
\hline
\end{tabular}
\end{table}

\section{Simulation-based evaluation} \label{sec:sim}
We implemented the CODA matching-based resource allocation in \textit{Python v.3.7.4} using the \textit{matching} 
library~\cite{wilde2020matching}. The script required to run the CODA model is available in the GitHub code repository\footnote{\url{https://github.com/SiNa88/CODA}\label{github}}.
We utilize the \textit{iFogSim} simulator~\cite{gupta2017ifogsim} to perform the evaluation on a simulated Cloud -- Fog environment.

\subsection{Resource setup} \label{ssec:simsetup}
Table~\ref{tbl:simconfg} displays the simulated Cloud -- Fog computing environment divided 
in three hierarchical tiers based on their computation, storage and networking capabilities.
We used the Phoronix test suite benchmark~\cite{phoronix} to measure the performance of each resource and then use it in the simulation. The measured computational $\CPU$ power of the resources is in the range $\SIrange{20000}{100000}{MIPS}$.
\subsubsection{Cloud data center} simulates instances equivalent to \texttt{m5a.8xlarge} of Amazon EC2, based on the 32-core AMD$^\circledR$ EPYC 7571 processor with a clock frequency of \SI{2.1}{\giga\hertz}. We select the \texttt{m5a.8xlarge} instance as it provides a good balance of computation, memory and network resources, suitable for executing data stream processing \cite{Li:2019}. 
\subsubsection{Fog-tier-2} simulates processing gateways (ISP GW) and cellular Base Transceiver Stations (BTS) available within Internet Service Providers (ISP) networks. We simulate the configuration based on the Alcatel-Lucent Ultimate Wireless Packet Core with 28-core Intel$^\circledR$ Xeon Platinum 8175 and base clock frequency of \SI{2.5}{\giga\hertz}~\cite{cloudlet}.
\subsubsection{Fog-tier-1} simulates resources co-located with the WiFi transceivers based on an eight-core Intel$^\circledR$ Core$^{(TM)}$ i7-7700 $\CPU$ at \SI{3.60}{\giga\hertz} equivalent configuration, widely used for data stream processing at the consumer premises~\cite{wificloudlet}.

\subsubsection{Interconnection network} simulates various Ethernet, wireless LAN, and 4G/LTE interfaces. We assume that gigabit switches interconnect the Cloud data centers and the Fog resources. As the Cloud interconnection network multiplexes the streaming traffic of multiple instances, the throughput to the Cloud data center is lower than to the Fog resources because of the shared bandwidth. Hence, we chose a bandwidth $\BW$ in the range $\SIrange{200}{1000}{\mega\bit\per\second}$ and a latency in the range $\SIrange{3}{100}{\milli\second}$ for interconnecting the Cloud and Fog resources. We derived these values from the maximum achievable bandwidth and the effective downlink throughput measured using the \texttt{iPerf3} tool~\cite{iperf3,zhang2017towards} (see Table~\ref{tbl:simconfg}).

\begin{table}[t]
\centering
\caption{Simulated Cloud -- Fog infrastructure.}
\label{tbl:simconfg}
\begin{tabular}{|c|c|c|c|}
\cline{2-4}
\multicolumn{1}{{c|}}{}&{\emph{Cloud}} &{\emph{Fog-tier-2}}&\multicolumn{1}{{c|}}{\emph{Fog-tier-1}}\\
\hline
\makecell{\emph{$\CPU$} {[MIPS]{$\cdot 10^{3}$}}} & 100 & \{80,75\} & \{20,30\}\\
\hline
\makecell{\emph{Memory} \SI{}{[\giga\byte]}} & 128 & \{64,32\} & \{8,16\}\\
\hline
\makecell{\emph{Storage} \SI{}{[\giga\byte]}} & 1200 & \{250,128\} & \{16,64\}\\
\hline
\makecell{\emph{BW} \SI{}{[\mega\bit\per\second]}} & \makecell{200} & \makecell{\{200,500\}} & \multicolumn{1}{{c|}}{1000}\\
\hline
\end{tabular}
\end{table}

\subsection{Experimental design} \label{ssec:experimntdesign}
We designed two sets of experiments according to the characteristics of the video stream processing application
.
\subsubsection{$\CPU$ experiment} varies the requirement in the range of $\{10000, 20000, 30000, 40000\}$ (MI) by bounding the data element to $\ELEM_{ui}[x] = \SI{10}{\mega\byte}$, which is the largest data element supported by the simulated communication protocol.
\subsubsection{Data experiment} varies the data element size $\ELEM_{ui}[x]$ transferred between microservices in the range $\{0.1, 1, 5, 10\}\ \SI{}{\mega\byte}$~
, with a $\CPU$ requirement of $15000$ MI.

\subsection{Performance metrics}\label{subsec:metrics}
We compare the performance of our CODA method against related works based on two metrics.
\subsubsection{Stream processing time} on the resources $\mu(\mathcal{A}) \subseteq \mathcal{R}$ is the completion time of the application $m_{\SINK}$ microservice:
\[C\left(\mathcal{A}, \mathcal{R}\right) = C\left(m_{\SINK}, \mathcal{R}\right),\]
where the completion time of a microservice $m_i$ is the maximum completion time of all its upstream microservices $C\left(m_u,\mathcal{R}\right)$ plus its microservice stream processing time $T\left(m_i, \STREAM_{ui}, r_j\right)$ on the allocated resource $r_j = \mu\left(m_i\right)$:
\begin{equation*}
\scriptsize
C\left(m_i, \mathcal{R}\right)=
\left\{
\arraycolsep=0pt
\begin{array}{ll}
T\left(m_{\SOURCE}, \SOURCE, r_j\right), & m_{\SOURCE} = m_i;\\
\max\limits_{\substack{\forall \left(m_u, m_i\right.,\\ \left. \STREAM_{ui}\right) \in \mathcal{E}}} \left\{C\left(m_u,\mathcal{R}\right)\right\} + T\left(m_i, \STREAM_{ui}, r_j\right), & m_{\SOURCE} \neq m_i,
\end{array}
\right.
\label{eq:compltime}
\end{equation*}

\subsubsection{Total streaming traffic} aggregates the traffic across all network channels. We define the streaming traffic on a network channel $l_{qj}$ as the ratio of all the data elements $\ELEM_{ui}[x]$ streaming between the resources $r_q=\mu\left(m_u\right)$ and $r_j=\mu\left(m_i\right)$ allocated to two interdependent microservices 
and the bandwidth $\BW_{qj}$ of a channel between the two resources:
\begin{equation*}
\mathit{Str\_Traf}\left(\mathcal{A}, \mathcal{R}\right) =\sum\limits_{\substack{\forall \left(m_u, m_i, \ELEM_{ui}\right) \in \mathcal{E}\\ \land\ l_{qj}\in \mathcal{L}}} \frac{\sum\limits_{x=1}^{\SIZE_{ui}} \left(\lambda_{ui}\cdot  \ELEM_{ui}[x]\right)}{\BW_{qj}}.
\label{eq:traffic}
\end{equation*}

\subsection{Related work comparison}
We conduct the performance comparisons against three state-of-the-art approaches divided in two categories.
\subsubsection{Cloud} uses only Cloud data centers for an application:
\paragraph{Heterogeneous Earliest Finish Time -- only Cloud (HEFT-oC)} deploys all microservices on the Cloud and selects the proper Cloud instances using a bottom ranking approach to optimize the stream processing time~\cite{topcuoglu2002performance}.

\subsubsection{Cloud and Fog} uses a combination of Cloud data centers and Fog resources.
\paragraph{Response Time Rate with Region Patterns (RTR-RP)}\cite{da2018latency} minimizes the stream processing time by analyzing the data flow patterns to deploy the microservices on the Fog resources that offer the shortest stream processing time. The Cloud data center only hosts the microservices that do not fit on the Fog devices due to resource and network requirements. 
\paragraph{CloudPath}\cite{mortazavi2017cloudpath} optimizes the stream processing time on a progression of Cloud data centers and Fog resources. CloudPath organizes the data centers in a multi-tier topology, and identifies first resources in the lowest tier (closest to the data $\SOURCE$) that meet the application requirements. If such resources are not available, it checks in the upper layers until it finds appropriate allocation resource.

\subsection{Simulation results}
Figures~\ref{sim:cpu} and~\ref{sim:data} illustrate the relation between the stream processing time and the total streaming traffic by increasing the computation and communication loads. 

\subsubsection{CPU experiment}
\paragraph{Stream processing time}
Figure~\ref{sim:cpu:time} shows that CODA reduces the stream processing time by $22$\%, $12$\%, and $15$\% compared to \mbox{RTR-RP}, HEFT-oC and CloudPath. The related methods allocate Cloud resources to the last microservices residing farther away from the data $\SOURCE$ in the application DAG, which explains this result. \mbox{RTR-RP} allocates the Cloud resource to the $\SINK$ microservice, which increases stream processing time, as the data needs to travel at least twice between the Cloud and the Fog-tier-1.

\paragraph{Total streaming traffic} Figure~\ref{sim:cpu:traffic} shows that CODA reduces the average streaming traffic by $5$\%, $8$\% and $7$\% compared to \mbox{RTR-RP}, HEFT-oC and CloudPath by allocating resources in the Fog-tier-2 instead of Cloud virtual machines. As the data element size does not vary during the simulation, the streaming traffic does not change for microservices with different $\CPU$ requirements.

\begin{figure}[t]
\centering
\begin{subfigure}[!ht]{0.48\textwidth}
\centering
\begin{tikzpicture}[scale=.55]
\centering
\begin{axis}[
        thick,
        ybar, axis on top,
        height=10cm, width=13cm,
        bar width=0.4cm,
        nodes near coords,
        every node near coord/.append style={rotate=90, anchor=west, font=\scriptsize},
        major grid style={draw=black},
        enlarge y limits={value=.1,upper},
        ymin=0, ymax=10,
        legend style={
            at={(0.5,-0.15)},
            anchor=north,
            legend columns=-1,
            /tikz/every even column/.append style={column sep=0.5cm}
        },
        visualization depends on=rawy\as\rawy, 
        nodes near coords={%
            \pgfmathprintnumber{\rawy}
        },
        ylabel={Stream processing time [\SI{}{\second}]},
        y label style = {xshift=-0.2em},
        yticklabel style = {xshift=-0.2em},
        symbolic x coords={
            10000,20000,30000,40000},
        xlabel={CPU Requirement [MI]},
        xtick=data,
        axis lines*=left,
        clip=false,
        area legend
    ]
    \addplot [color=red, fill=red!30!] coordinates {
        (10000,6.302)(20000,6.797)(30000,7.605)(40000,7.038)
        };
    \addplot [color=orange, fill=orange!30!] coordinates {
        (10000,7.7)(20000,8.323)(30000,9.630)(40000,9.7)
        };
    \addplot [color =blue, fill=blue!30!] coordinates {
        (10000,7.519)(20000,7.520)(30000,8.477)(40000,8.478)
       };
    \addplot [color=green, fill=green!30!] coordinates {
        (10000,7.630)(20000,8.125)(30000,8.933)(40000,8.968)
        };
    \legend{CODA,  RTR-RP, HEFT-oC, CloudPath}
\end{axis}
\end{tikzpicture}
\caption{}
\label{sim:cpu:time}
\end{subfigure}
\hfill
\begin{subfigure}[t]{0.48\textwidth}
\centering
\begin{tikzpicture}[scale=.55]
\begin{axis}[
    thick,
    ybar, axis on top,
    height=10cm, width=13cm,
    bar width=0.4cm,
    nodes near coords,
    every node near coord/.append style={rotate=90, anchor=west},
    major grid style={draw=black},
    enlarge y limits={value=.1,upper},
    ymin=0, ymax=21000,
    legend style={
            at={(0.5,-0.15)},
            anchor=north,
            legend columns=-1,
           /tikz/every even column/.append style={column sep=0.5cm}
    },
    visualization depends on=rawy\as\rawy, 
    nodes near coords={%
        \pgfmathprintnumber{\rawy}
    },
    ylabel={Total streaming traffic},
    y label style = {xshift=-0.2em},
    yticklabel style = {xshift=-0.2em},
    symbolic x coords={10000,20000,30000,40000},
    xtick=data,
    axis lines*=left,
    clip=false,
    area legend,
    xlabel={CPU Requirement [MI]},
    ]	
\addplot[
    color=red, fill=red!30!,
    ]
    coordinates {
   (10000,19244)(20000,19210)(30000,19244)(40000,19244)
   };
\addplot[
    color=orange, fill=orange!30!,
    ]
    coordinates {
    (10000,20264)(20000,20264)(30000,20264)(40000,20264)
    };
\addplot[
    color=blue, fill=blue!30!,
    ]
    coordinates {
    (10000,20867)(20000,20867)(30000,20867)(40000,20867)
    };
\addplot[
    color=green, fill=green!30!,
    ]
    coordinates {
    (10000,20649)(20000,20649)(30000,20649)(40000,20649)
    };
    \legend{CODA, RTR-RP, HEFT-oC, CloudPath}
\end{axis}
\end{tikzpicture}
\caption{}
\label{sim:cpu:traffic}
\end{subfigure}
\caption{Simulated video application stream processing time and total traffic comparison for different CPU requirements.}
\label{sim:cpu}
\end{figure}

\subsubsection{Data experiment}
\paragraph{Stream processing time}
Figure~\ref{sim:data:time} shows that CODA reduces the stream processing time by $8$\%, $9.7$\% and $11$\% compared to \mbox{RTR-RP}, HEFT-oC and CloudPath for different data element sizes. Unlike the other approaches, CODA considers the network bandwidth and the data element in its microservice-side ranking to find matches that reduce the streaming traffic, and consequently the processing time.

\paragraph{Total streaming traffic}
Figure~\ref{sim:data:traffic} shows that HEFT-oC and CloudPath generate higher streaming traffic than CODA and \mbox{RTR-RP} for small data elements. As the data element increases, the streaming traffic gradually saturates the network channels, and the related approaches perform almost equally well. CODA reduces the streaming traffic up to $2.3$\% compared to the related methods by considering the data element in its resource-side ranking.

\begin{figure}[t]
\begin{subfigure}[!ht]{0.48\textwidth}
\centering
\begin{tikzpicture}[scale=.55]
\begin{axis}[thick,
        ybar, axis on top,
        height=10cm, width=13cm,
        bar width=0.4cm,
        nodes near coords,
        every node near coord/.append style={rotate=90, anchor=west},
        major grid style={draw=black},
        enlarge y limits={value=.1,upper},
        ymin=0, ymax=7.500,
        legend style={
            at={(0.5,-0.15)},
            anchor=north,
            legend columns=-1,
            /tikz/every even column/.append style={column sep=0.5cm}
        },
        visualization depends on=rawy\as\rawy, 
        nodes near coords={%
            \pgfmathprintnumber{\rawy}
        },
        ylabel={Stream processing time [\SI{}{\second}]},
        y label style = {xshift=-0.2em},
        yticklabel style = {xshift=-0.2em},
        symbolic x coords={0.1,1,5,10},
        xtick=data,
        axis lines*=left,
    clip=false,
    area legend,
    xlabel={Data element [\SI{}{\mega\byte}]},
    ]
\addplot[
    color=red, fill=red!30!,
    ]
    coordinates {
    (0.1,4.910)(1,5.210)(5,5.916)(10,6.302)
    };
\addplot[
    color=orange, fill=orange!30!,
    ]
    coordinates {
    (0.1,5.040)(1,5.522)(5,6.107)(10,7.720)
    };
\addplot[
    color=blue, fill=blue!30!
    ]
    coordinates {
    (0.1,5.023)(1,5.418)(5,6.876)(10,7.519)
    };
\addplot[
    color=green, fill=green!30!,
    ]
    coordinates {
    (0.1,5.078)(1,5.529)(5,6.987)(10,7.630)
    };
\legend{CODA, RTR-RP, HEFT-oC,CloudPath}
\end{axis}
\end{tikzpicture}
\caption{}
\label{sim:data:time}
\end{subfigure}
\hfill
\begin{subfigure}[!ht]{0.48\textwidth}
\centering
\begin{tikzpicture}[scale=.55]
\begin{axis}[
    thick,
    ybar, axis on top,
    height=10cm, width=13cm,
    bar width=0.4cm,
    nodes near coords,
    every node near coord/.append style={rotate=90, anchor=west},
    major grid style={draw=black},
    enlarge y limits={value=.1,upper},
    ymin=0, ymax=21000,
    legend style={
            at={(0.5,-0.15)},
            anchor=north,
            legend columns=-1,
           /tikz/every even column/.append style={column sep=0.5cm}
    },
    visualization depends on=rawy\as\rawy, 
    nodes near coords={%
        \pgfmathprintnumber{\rawy}
    },
    ylabel={Total streaming traffic},
    y label style = {xshift=-0.2em},
    yticklabel style = {xshift=-0.2em},
    symbolic x coords={0.1,1,5,10},
    xtick=data,
    axis lines*=left,
    clip=false,
    area legend,
    xlabel={Data element [\SI{}{\mega\byte}]},
    ]	
\addplot[
    color=red, fill=red!30!,
    ]
    coordinates {
    (0.1,19244)(1,18758)(5,17546)(10,14547)
   };
\addplot[
    color=orange, fill=orange!30!,
    ]
    coordinates {
    (0.1,20264)(1,19773)(5,17211)(10,13255)
    };
\addplot[
    color=blue, fill=blue!30!,
    ]
    coordinates {
    (0.1,20867)(1,20383)(5,17211)(10,13255)
    };
\addplot[
    color=green, fill=green!30!,
    ]
    coordinates {
    (0.1,20649)(1,20159)(5,17211)(10,13251)
    };
    \legend{CODA, RTR-RP, HEFT-oC, CloudPath}
\end{axis}
\end{tikzpicture}
\caption{}
\label{sim:data:traffic}
\end{subfigure}
\caption{Simulated video application stream processing time and total traffic for different data element sizes.}
\label{sim:data}
\end{figure}

\section{Real testbed evaluation}\label{sec:testbed}
To validate the simulation results, we analyze in this section the CODA performance on a real experimental testbed.

\subsection{Carinthian Computing Continuum} \label{ssec:restestbedsetup}
\begin{figure}[t]
    \centering
    \includegraphics[width=\columnwidth]{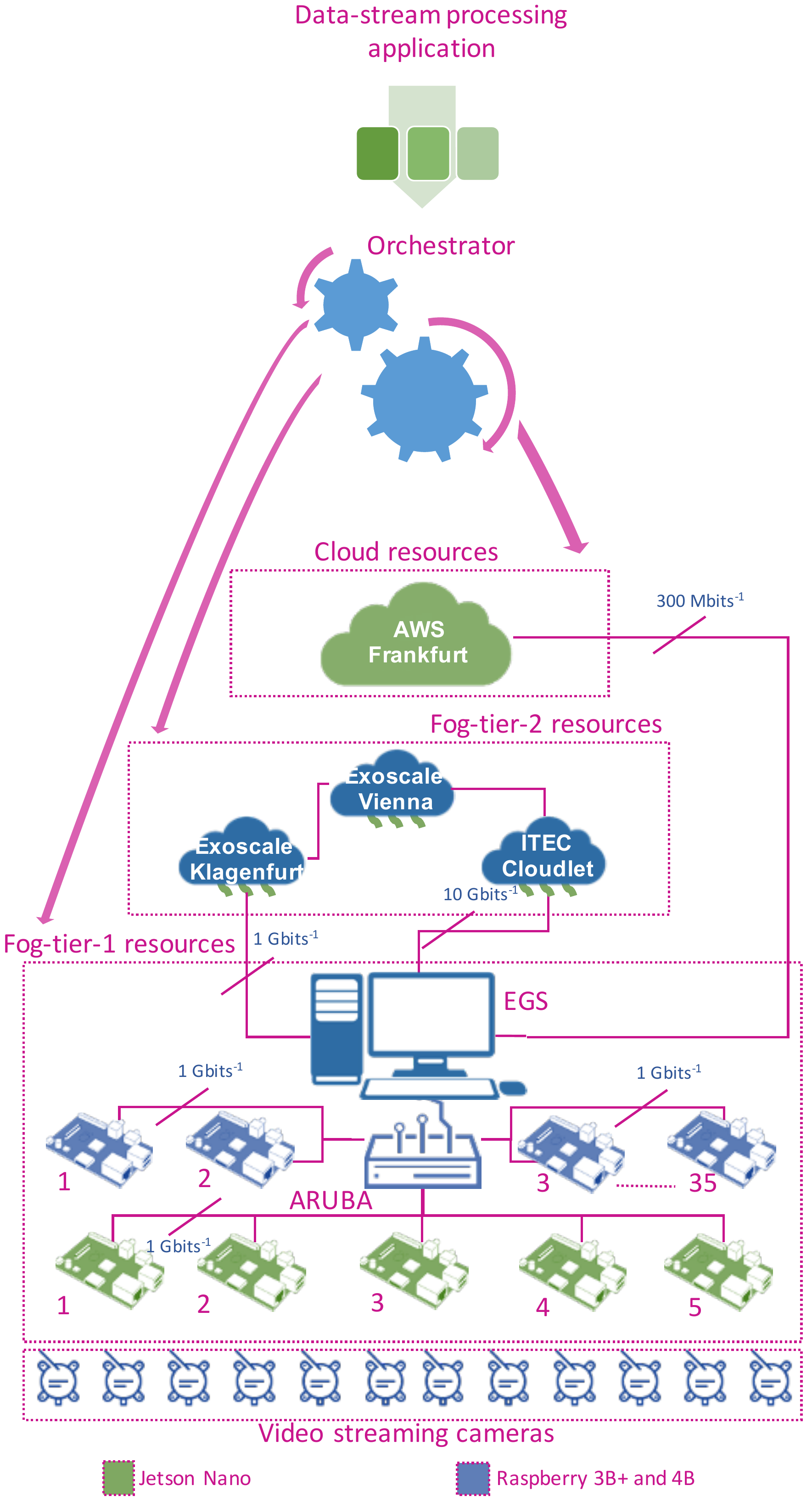}
    \caption{The $C^3$ testbed architecture.}
    \label{fig:c3}
\end{figure}

We deployed a real testbed at the University of Klagenfurt named \emph{Carinthian Computing Continuum} ($C^3$)~\cite{c3} that aggregates heterogeneous resources in three hierarchical categories~\cite{kimovski2021cloud}, as depicted in Figure~\ref{fig:c3}.

\subsubsection{Cloud data center} consists of virtualized instances provisioned on-demand from the Amazon Web Services (AWS), located at the geographically closest European data center in Frankfurt (Germany). We selected the \texttt{m5a.xlarge} general purpose instance powered by AMD EPYC 7000 processors at \SI{2.5}{\giga\hertz} and up to \SI{10}{\giga\bit\per\second} network bandwidth as the most suitable instance for our case study.

\subsubsection{Fog-tier-2} comprises resources from two providers, Exoscale~\cite{exoscale} and University of Klagenfurt, thanks to their low round-trip communication latency ($\leq$ \SI{7}{\milli\second}) and high bandwidth ($\leq$ \SI{10}{\giga\bit\per\second}). University of Klagenfurt provides a private Cloud infrastructure (PCI) using \texttt{OpenStack v13.0} and \texttt{Ceph v12.2} with support for block and S3-compatible object storage. The computing optimized instances are of type \texttt{large} running Ubuntu 18.04 LTS, as described in Table~\ref{tbl:testbed}.

\subsubsection{Fog-tier-1} comprises five NVIDIA Jetson Nano (NJN), three Raspberry Pi-3 B+ (RPi3B+), and $32$ Raspberry Pi-4 single-board computers (RPi4). We installed \texttt{Raspberry Pi OS} (version 2020-05-27)~
on all RPis and \texttt{Linux for Tegra (L4T)} operating system~
on NJN resources. A managed layer-3 HP Aruba switch interconnects the Fog-tier-1 resources. The switch has 48 \SI{1}{\giga\bit\per\second} ports, a latency of \SI{3.8}{\micro\second} and an aggregate data transfer rate of \SI{104}{\giga\bit\per\second}. 
The Fog-tier-1 has a Fog/Edge Gateway System (EGS) as the entry point to the other resources available in this tier. The EGS has a twelve-core AMD Ryzen Threadripper 2920X processor at \SI{3.5}{\giga\hertz} and \SI{32}{\giga\byte} of RAM, running Ubuntu 18.04 LTS. It supports \SI{1}{\giga\bit\per\second} Ethernet and dual band PCIe WiFi 5 (802.11ac) network connections.

\begin{table}[t]
\centering
\caption{The $C^3$ testbed configuration.}
\label{tbl:testbed}
\resizebox{\columnwidth}{!}{
\begin{tabular}{|@{}c@{}|@{}c@{}|@{}c@{}|@{}c@{}|@{}c@{}|@{}c@{}|@{}c@{}|}
\cline{2-7}
\multicolumn{1}{{c|}}{}&{\emph{Cloud}} &\multicolumn{1}{{c|}}{\emph{Fog-tier-2}}&\multicolumn{4}{{c|}}{\emph{Fog-tier-1}}\\
\hline
\makecell{\textit{Instance / Device}}&\makecell{AWS \texttt{m5a.xlarge}} & \makecell{Exoscale \texttt{Large} \\ PCI \texttt{Large}}&\makecell{EGS}&\makecell{NJN}&\makecell{RPi4}&\makecell{RPi3B+} \\
\hline 
\makecell{\textit{CPU type}}&\makecell{AMD EPYC 7000} & \makecell{Intel Xeon \\ Platinum 8180}&\makecell{AMD Ryzen \\  2920}&\makecell{Tegra X1 and \\ ARM Cortex A57}&\makecell{ARM \\Cortex 72}&\makecell{ARM \\Cortex 53} \\
\hline
\makecell{\emph{CPU clock} {[\SI{}{\giga\hertz}]}} & 2.5 & 3.6& 3.5 & 1.43 & 1.5 & 1.4\\
\hline
\makecell{\emph{Memory} \SI{}{[\giga\byte]}} & 32 & 8 & 32 & 4 & 4 & 1\\
\hline
\makecell{\emph{Storage} \SI{}{[\giga\byte]}} & 1,000 &256& 1,000 &64 & 64 & 64\\
\hline
\makecell{\emph{$\BW$} \SI{}{[\mega\bit\per\second]}} &\makecell{27}&\makecell{65}& \makecell{813} & \makecell{450}& \makecell{800}& \makecell{330}\\
\hline
\end{tabular}
}
\end{table}
We installed a Docker engine 19.03 
on all resources and containerized each microservice in Ubuntu 18.10 Docker official image. The minimal scripts to create and run the containerized microservices on the resources is available in the GitHub code repository\textsuperscript{\ref{github}}.

\subsection{Experimental design} \label{ssec:tbexperimntdesign}
We evaluated CODA compared to the related HEFT-oC, \mbox{RTR-RP} and CloudPath methods using the video stream processing application for traffic sign classification, described in Section~\ref{sec:casestudy}.
We processed a raw video stream of \SI{9}{\second} and \SI{45}{\mega\byte} in size that includes the traffic signs
. 
We designed two sets of experiments according to the application characteristics.

\subsubsection{$\CPU$ experiments} investigate the impact of $\CPU$ requirements for transcoding the raw video segment. We considered four encoding and transcoding bit rates of $\left\{200, 1500, 3000, 6500, 20000\right\}$ \SI{}{\kilo\bit\per\second}, corresponding to resolutions of $\left\{180, 576, 720, 1440, 2160\right\}$ pixels. We further considered two machine learning models with $70\%$ and $90\%$ accuracy for the inference microservices with different $\CPU$ requirements. We fixed the size of the data element to $\SI{2560}{\kilo\byte}$.

\subsubsection{Data experiments} compare the different methods using data element sizes in the range: $\ELEM_{ui}[x] \in \left\{35, 300, 420, 1350, 2560 \right\}\SI{}{\kilo\byte}$, which correspond to different video frame sizes obtained by using five different qualities. We fixed the video resolution to $2160p$.

\subsection{Real-world testbed results}
\subsubsection{CPU experiments}
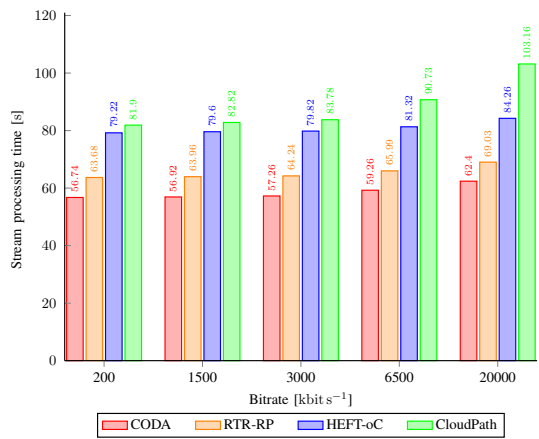
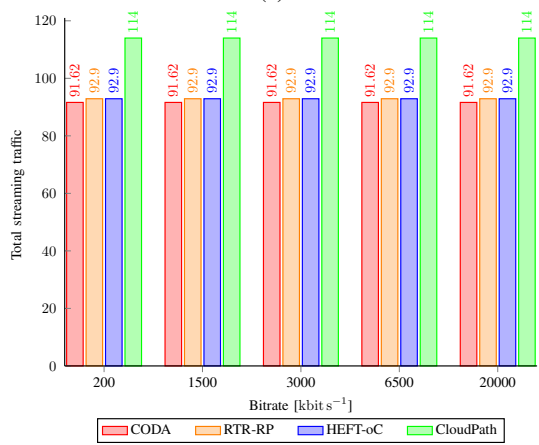
\begin{figure}[!t]
\centering
\begin{subfigure}[t]{0.48\textwidth}
\centering
\begin{tikzpicture}[scale=.55]
  \centering
  \begin{axis}[
        thick,
        ybar, axis on top,
        height=10cm, width=13cm,
        bar width=0.4cm,
        nodes near coords,
        every node near coord/.append style={rotate=90, anchor=west, font=\scriptsize},
        major grid style={draw=black},
        enlarge y limits={value=.1,upper},
        ymin=0, ymax=110,
        legend style={
            at={(0.5,-0.15)},
            anchor=north,
            legend columns=-1,
            /tikz/every even column/.append style={column sep=0.5cm}
        },
        visualization depends on=rawy\as\rawy, 
        nodes near coords={%
            \pgfmathprintnumber{\rawy}
        },
        ylabel={Stream processing time [\SI{}{\second}]},
        y label style = {xshift=-0.2em},
        yticklabel style = {xshift=-0.2em},
        symbolic x coords={200,1500,3000,6500,20000},
        xlabel={Bitrate [\SI{}{\kilo\bit\per\second}]},
        xtick=data,
        axis lines*=left,
        clip=false,
        area legend
    ]
    \addplot [color=red, fill=red!30!] coordinates {
        (200,56.7418)
        (1500, 56.9218) 
        (3000, 57.2618)
        (6500, 59.2618)
        (20000, 62.4018)
        };
    \addplot [color=orange, fill=orange!30!] coordinates {
        (200, 63.6847)
        (1500, 63.9647)
        (3000, 64.2447)
        (6500, 65.9947)
        (20000, 69.0347)
        };
    \addplot [color =blue, fill=blue!30!] coordinates {
        (200, 79.2164)
        (1500, 79.5964)
        (3000, 79.8164)
        (6500, 81.3164)
        (20000, 84.2564)
       };
    \addplot [color=green, fill=green!30!] coordinates {
        (200, 81.9013)
        (1500, 82.8213) 
        (3000, 83.7813)
        (6500,  90.7313)
        (20000, 103.1613)
        };
    \legend{CODA,  RTR-RP, HEFT-oC, CloudPath}
\end{axis}
\end{tikzpicture}
\caption{}
\label{testbed:cpu:time}
\end{subfigure}
\hfill
\begin{subfigure}[t]{0.48\textwidth}
\centering
\begin{tikzpicture}[scale=.55]
\begin{axis}[
    thick,
    ybar, axis on top,
    height=10cm, width=13cm,
    bar width=0.4cm,
    nodes near coords,
    every node near coord/.append style={rotate=90, anchor=west},
    major grid style={draw=black},
    enlarge y limits={value=.1,upper},
    ymin=0, ymax=110,
    legend style={
            at={(0.5,-0.15)},
            anchor=north,
            legend columns=-1,
           /tikz/every even column/.append style={column sep=0.5cm}
    },
    visualization depends on=rawy\as\rawy, 
    nodes near coords={%
        \pgfmathprintnumber{\rawy}
    },
    ylabel={Total streaming traffic},
    y label style = {xshift=-0.2em},
    yticklabel style = {xshift=-0.2em},
    symbolic x coords={200,1500,3000,6500,20000},
    xtick=data,
    axis lines*=left,
    clip=false,
    area legend,
    xlabel={Bitrate [\SI{}{\kilo\bit\per\second}]},
    ]	
\addplot[
    color=red, fill=red!30!,
    ]
    coordinates {
        (200, 91.62)
        (1500, 91.62) 
        (3000, 91.62)
        (6500,  91.62)
        (20000, 91.62)
   };
\addplot[
    color=orange, fill=orange!30!,
    ]
    coordinates {
        (200, 92.9)
        (1500, 92.9) 
        (3000, 92.9)
        (6500,  92.9)
        (20000, 92.9)
    };
\addplot[
    color=blue, fill=blue!30!,
    ]
    coordinates {
        (200, 92.9)
        (1500, 92.9) 
        (3000, 92.9)
        (6500,  92.9)
        (20000, 92.9)
    };
\addplot[
    color=green, fill=green!30!,
    ]
    coordinates {
        (200, 114)
        (1500, 114) 
        (3000, 114)
        (6500,  114)
        (20000, 114)
    };
    \legend{CODA, RTR-RP, HEFT-oC, CloudPath}
\end{axis}
\end{tikzpicture}
\caption{}
\label{testbed:cpu:traffic}
\end{subfigure}
\caption{Real video application stream processing time and total traffic for different bitrates.}
\label{testbed:cpu}
\end{figure}

\paragraph{Stream processing time}
Figure~\ref{testbed:cpu:time} shows that CODA reduces the stream processing time by $11$\%, $28$\% and $33$\% compared to \mbox{RTR-RP}, HEFT-oC and CloudPath. CODA performs the video encoding on Fog-tier-1 resources (NJN and RPi4), and enables video transcoding and high-accuracy inference on the EGS. The \mbox{RTR-RP} and CloudPath methods tend to allocate resources from Fog-tier-2 (i.e. Exoscale \texttt{Large}, PCI \texttt{Large}) with higher communication latency and similar computing performance to EGS. Lastly, HEFT-oC utilizes the AWS \texttt{m5a.xlarge} instances with high computing performance but limited measured bandwidth of \SI{27}{\mega\bit\per\second}.    

\paragraph{Total streaming traffic} Figure~\ref{testbed:cpu:traffic} shows that
all methods except CloudPath exhibit similar performance as the data element size increases.
CloudPath introduces 
up to $16$\% higher streaming traffic than CODA because it allocates Fog-tier-1 and Cloud resources, which require the data to traverse more network channels from the $\SOURCE$.
CODA deploys encoding microservices onto Fog-tier-2 resources closer to the data $\SOURCE$ and hence, the raw video stream traverses less network channels with lower streaming traffic.


\subsubsection{Data experiments}

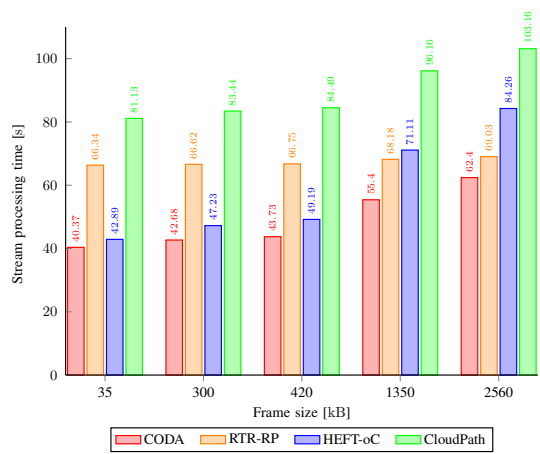
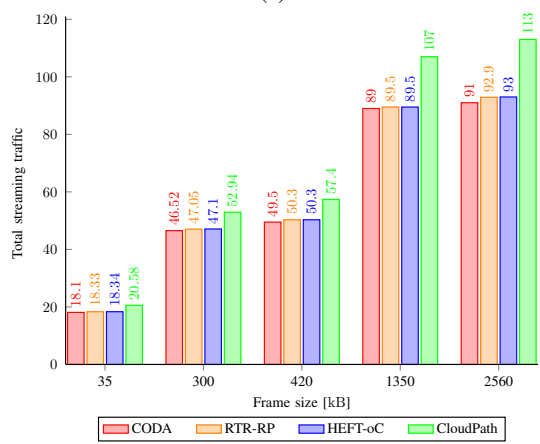
\begin{figure}[!t]
\centering
\begin{subfigure}[!ht]{0.48\textwidth}
\centering
\begin{tikzpicture}[scale=.55]
  \centering
  \begin{axis}[
        thick,
        ybar, axis on top,
        height=10cm, width=13cm,
        bar width=0.4cm,
        nodes near coords,
        every node near coord/.append style={rotate=90, anchor=west, font=\scriptsize},
        major grid style={draw=black},
        enlarge y limits={value=.1,upper},
        ymin=0, ymax=100,
        legend style={
            at={(0.5,-0.15)},
            anchor=north,
            legend columns=-1,
            /tikz/every even column/.append style={column sep=0.3cm}
        },
        visualization depends on=rawy\as\rawy, 
        nodes near coords={%
            \pgfmathprintnumber{\rawy}
        },
        ylabel={Stream processing time [\SI{}{\second}]},
        y label style = {xshift=-0.2em},
        yticklabel style = {xshift=-0.2em},
        symbolic x coords={35, 300, 420, 1350, 2560},
        xlabel={Frame size [\SI{}{\kilo\byte}]},
        xtick=data,
        axis lines*=left,
        clip=false,
        area legend
    ]
    \addplot [color=red, fill=red!30!] coordinates {
        (35, 40.368) 
        (300, 42.6805)
        (420,43.7276)
        (1350, 55.4017)
        (2560, 62.4018)
        };
    \addplot [color=orange, fill=orange!30!] coordinates {
        (35, 66.3367) 
        (300, 66.6199)
        (420, 66.7481)
        (1350, 68.1776)
        (2560, 69.0347)
        };
    \addplot [color =blue, fill=blue!30!] coordinates {
        (35, 42.8868) 
        (300, 47.2285)
        (420, 49.1946)
        (1350, 71.1133)
        (2560, 84.2564)
       };
    \addplot [color=green, fill=green!30!] coordinates {
        (35, 81.1275) 
        (300, 83.44)
        (420, 84.4871)
        (1350, 96.1612)
        (2560, 103.1613)
        };
    \legend{CODA, RTR-RP, HEFT-oC, CloudPath}
  \end{axis}
\end{tikzpicture}
\caption{}
\label{testbed:data:time}
\end{subfigure}
\hfill
\begin{subfigure}[!ht]{0.48\textwidth}
\centering
\begin{tikzpicture}[scale=.55]
\begin{axis}[
    thick,
    ybar, axis on top,
    height=10cm, width=13cm,
    bar width=0.4cm,
    nodes near coords,
    every node near coord/.append style={rotate=90, anchor=west},
    major grid style={draw=black},
    enlarge y limits={value=.1,upper},
    ymin=0, ymax=110,
    legend style={
            at={(0.5,-0.15)},
            anchor=north,
            legend columns=-1,
           /tikz/every even column/.append style={column sep=0.5cm}
    },
    visualization depends on=rawy\as\rawy, 
    nodes near coords={%
        \pgfmathprintnumber{\rawy}
    },
    ylabel={Total streaming traffic},
    y label style = {xshift=-0.2em},
    yticklabel style = {xshift=-0.2em},
    symbolic x coords={35, 300, 420, 1350, 2560},
    xtick=data,
    axis lines*=left,
    clip=false,
    area legend,
    xlabel={Frame size [\SI{}{\kilo\byte}]},
    ]	
\addplot[
    color=red, fill=red!30!,
    ]
    coordinates {
        (35,  18.1) 
        (300, 46.52)
        (420, 49.5)
        (1350, 89.0)
        (2560, 91.0)
   };
\addplot[
    color=orange, fill=orange!30!,
    ]
    coordinates {
        (35,  18.33) 
        (300, 47.05)
        (420, 50.3)
        (1350, 89.5)
        (2560, 92.9)
    };
\addplot[
    color=blue, fill=blue!30!,
    ]
    coordinates {
        (35, 18.34) 
        (300, 47.1)
        (420, 50.3)
        (1350, 89.5)
        (2560, 93)
    };
\addplot[
    color=green, fill=green!30!,
    ]
    coordinates {
        (35,  20.58) 
        (300, 52.94)
        (420, 57.4)
        (1350, 107.0)
        (2560, 113.0)
    };
    \legend{CODA, RTR-RP, HEFT-oC, CloudPath}
\end{axis}
\end{tikzpicture}
\caption{}
\label{testbed:data:traffic}
\end{subfigure}
\caption{Real video application stream processing time and total traffic for different frame sizes.}
\label{testbed:dataelement}
\end{figure}

\paragraph{Stream processing time} Figure~\ref{testbed:data:time}
shows that CODA outperforms \mbox{RTR-RP}, HEFT-oC, and CloudPath by $37$\%, $16$\%, and $45$\% on average by deploying microservices on resources closer to the application data $\SOURCE$ and $\SINK$. CODA reduces the stream processing time by performing the video encoding on the NJN and RPi4 devices. This considerably reduces the streaming traffic, as the encoded video is significantly smaller for the same data element size. HEFT-oC generates the highest stream processing time by transferring a raw video stream from the $\SOURCE$ to the \texttt{m5a.xlarge} instance in the AWS data center. Finally, RTR-RP and CloudPath show similar performance because they use more distant Fog-tier-2 resources for encoding, despite performing the machine learning training on the computationally-efficient EGS device.

\paragraph{Total streaming traffic} Figure~\ref{testbed:data:traffic} shows that CODA reduces the streaming traffic by $1.3$\%, $1.4$\% and $16$\% on average compared to \mbox{RTR-RP}, HEFT-oC and CloudPath. CODA reduces the streaming traffic by allocating microservices to resources in the Fog-tier-2 layer with lower stream processing time. In contrast, CloudPath requires the data stream to traverse more network channels towards the data $\SINK$, which generates higher streaming traffic.


\subsubsection{Conclusion} The real testbed evaluation confirms the simulation. Surprisingly, the benefits of CODA to the stream processing application are even higher compared to the three related methods due to the higher latency and lower bandwidth difference between the Fog-tier-1, Fog-tier-2 resources and the Cloud instances within the $C^3$ testbed.

\section{Conclusions and future work} \label{sec:concl}
We introduced CODA, a novel approach for allocating heterogeneous Cloud -- Fog computing resources to data stream processing applications, described as DAGs. CODA applies a two-sided stable matching model that enables many-to-one assignment of application microservices to resources based on specific ranking strategies.
The microservices rank the continuum resources based on their microservice stream processing time. On the other side, resources rank the stream processing microservices based on their residual bandwidth. A two-sided stable matching model assigns microservices to resources based on their mutual preferences, aiming to optimize the complete stream processing time on the application side and the total streaming traffic on the resource side.
We evaluated CODA based on a video stream processing application for traffic sign classification using comprehensive simulation combined with a real Cloud -- Fog experimental testbed deployment. The results demonstrate that CODA achieves \SIrange{11}{45}{}\% lower stream processing times and \SIrange{1.3}{20}{}\% lower streaming traffic than three state-of-the-art approaches. 
In the future, we plan to further improve our results 
by analyzing Nash equilibrium 
while processing the data streams in the Cloud -- Fog computing continuum.

\section*{Acknowledgement}
This work received support from the DataCloud project funded by the European Commission under the Horizon 2020 Programme (grant number 101016835) and the K{\"a}rntner Fog 5G Playground project funded by Carinthian Agency for Investment Promotion and Public Shareholding.

\bibliography{ref}

\begin{thebibliography}{10}

\bibitem{lai2018optimal}
Phu Lai, Qiang He, Mohamed Abdelrazek, Feifei Chen, John Hosking, John Grundy,
  and Yun Yang.
\newblock Optimal edge user allocation in edge computing with variable sized
  vector bin packing.
\newblock In {\em International Conference on Service-Oriented Computing},
  pages 230--245. Springer, 2018.

\bibitem{mortazavi2017cloudpath}
Seyed~Hossein Mortazavi, Mohammad Salehe, Carolina~Simoes Gomes, Caleb
  Phillips, and Eyal de~Lara.
\newblock Cloudpath: A multi-tier cloud computing framework.
\newblock In {\em Proceedings of the Second ACM/IEEE Symposium on Edge
  Computing}, pages 1--13, 2017.

\bibitem{aral2019addressing}
Atakan Aral, Ivona Brandic, Rafael~Brundo Uriarte, Rocco De~Nicola, and
  Vincenzo Scoca.
\newblock Addressing application latency requirements through edge scheduling.
\newblock {\em Journal of Grid Computing}, pages 1--22, 2019.

\bibitem{marin2017we}
Eva Mar{\'\i}n-Tordera, Xavi Masip-Bruin, Jordi Garc{\'\i}a-Almi{\~n}ana,
  Admela Jukan, Guang-Jie Ren, and Jiafeng Zhu.
\newblock Do we all really know what a fog node is? current trends towards an
  open definition.
\newblock {\em Computer Communications}, 109:117--130, 2017.

\bibitem{kekki2018mec}
Sami Kekki, Walter Featherstone, Yonggang Fang, Pekka Kuure, Alice Li, Anurag
  Ranjan, Debashish Purkayastha, Feng Jiangping, Danny Frydman, Gianluca Verin,
  et~al.
\newblock Mec in 5g networks.
\newblock {\em Sophia Antipolis, France, ETSI, White Paper}, 2018.

\bibitem{de2018distributed}
Marcos~Dias de~Assuncao, Alexandre da~Silva~Veith, and Rajkumar Buyya.
\newblock Distributed data stream processing and edge computing: A survey on
  resource elasticity and future directions.
\newblock {\em Journal of Network and Computer Applications}, 103:1--17, 2018.

\bibitem{abedin2018resource}
Sarder~Fakhrul Abedin, Md~Golam~Rabiul Alam, SM~Ahsan Kazmi, Nguyen~H Tran,
  Dusit Niyato, and Choong~Seon Hong.
\newblock Resource allocation for ultra-reliable and enhanced mobile broadband
  iot applications in fog network.
\newblock {\em IEEE Transactions on Communications}, 67(1):489--502, 2018.

\bibitem{gupta2017ifogsim}
Harshit Gupta, Amir Vahid~Dastjerdi, Soumya~K Ghosh, and Rajkumar Buyya.
\newblock ifogsim: A toolkit for modeling and simulation of resource management
  techniques in the internet of things, edge and fog computing environments.
\newblock {\em Software: Practice and Experience}, 47(9):1275--1296, 2017.

\bibitem{sharghivand2018qos}
Nafiseh Sharghivand, Farnaz Derakhshan, and Lena Mashayekhy.
\newblock Qos-aware matching of edge computing services to internet of things.
\newblock In {\em 2018 IEEE 37th International Performance Computing and
  Communications Conference (IPCCC)}, pages 1--8. IEEE, 2018.

\bibitem{cai2018response}
Xinchen Cai, Hongyu Kuang, Hao Hu, Wei Song, and Jian L{\"u}.
\newblock Response time aware operator placement for complex event processing
  in edge computing.
\newblock In {\em International Conference on Service-Oriented Computing},
  pages 264--278. Springer, 2018.

\bibitem{da2018latency}
Alexandre da~Silva~Veith, Marcos~Dias de~Assun{\c{c}}ao, and Laurent Lefevre.
\newblock Latency-aware placement of data stream analytics on edge computing.
\newblock In {\em International Conference on Service-Oriented Computing},
  pages 215--229. Springer, 2018.

\bibitem{dautov2020stream}
Rustem Dautov and Salvatore Distefano.
\newblock Stream processing on clustered edge devices.
\newblock {\em IEEE Transactions on Cloud Computing}, 2020.

\bibitem{zamani2017deadline}
Ali~Reza Zamani, Mengsong Zou, Javier Diaz-Montes, Ioan Petri, Omer~Farooq
  Rana, Ashiq Anjum, and Manish Parashar.
\newblock Deadline constrained video analysis via in-transit computational
  environments.
\newblock {\em IEEE Transactions on Services Computing}, 2017.

\bibitem{ehsanpour2019efficient}
Mahsa Ehsanpour, Siavash Bayat, and Ali Mohammad~Afshin Hemmatyar.
\newblock An efficient and social-aware distributed in-network caching scheme
  in named data networks using matching theory.
\newblock {\em Computer Networks}, 158:175--183, 2019.

\bibitem{zhao2016pricing}
Tianchu Zhao, Sheng Zhou, Xueying Guo, Yun Zhao, and Zhisheng Niu.
\newblock Pricing policy and computational resource provisioning for
  delay-aware mobile edge computing.
\newblock In {\em ICCC}, pages 1--6, 2016.

\bibitem{mao2017draps}
Ying Mao, Jenna Oak, Anthony Pompili, Daniel Beer, Tao Han, and Peizhao Hu.
\newblock Draps: Dynamic and resource-aware placement scheme for docker
  containers in a heterogeneous cluster.
\newblock In {\em 2017 IEEE 36th International Performance Computing and
  Communications Conference (IPCCC)}, pages 1--8. IEEE, 2017.

\bibitem{birkenheuer2008virtual}
Georg Birkenheuer, Andr{\'e} Brinkmann, Hubert D{\"o}mer, Sascha Effert,
  Christoph Konersmann, Oliver Nieh{\"o}rster, and Jens Simon.
\newblock Virtual supercomputer for hpc and htc.
\newblock {\em Joint workshop of the GI / ITG specialist groups on "der GI/ITG
  Fachgruppen Betriebssysteme und KuVS"}, pages 37--50, 2008.

\bibitem{VirtualizationViaContainers}
Donald Firesmith.
\newblock Virtualization via containers, 2017.
\newblock
  \url{https://insights.sei.cmu.edu/sei_blog/2017/09/virtualization-via-containers.html}.

\bibitem{DEMAIO2020171}
Vincenzo {De Maio} and Dragi Kimovski.
\newblock Multi-objective scheduling of extreme data scientific workflows in
  fog.
\newblock {\em Future Generation Computer Systems}, 106:171 -- 184, 2020.

\bibitem{mehran2019mapo}
Narges Mehran, Dragi Kimovski, and Radu Prodan.
\newblock Mapo: A multi-objective model for iot application placement in a fog
  environment.
\newblock In {\em Proceedings of the 9th International Conference on the
  Internet of Things}, pages 1--8, 2019.

\bibitem{bayat2016matching}
Siavash Bayat, Yonghui Li, Lingyang Song, and Zhu Han.
\newblock Matching theory: Applications in wireless communications.
\newblock {\em IEEE Signal Processing Magazine}, 33(6):103--122, 2016.

\bibitem{gale1962college}
David Gale and Lloyd~S Shapley.
\newblock College admissions and the stability of marriage.
\newblock {\em The American Mathematical Monthly}, 69(1):9--15, 1962.

\bibitem{diebold2014course}
Franz Diebold, Haris Aziz, Martin Bichler, Florian Matthes, and Alexander
  Schneider.
\newblock Course allocation via stable matching.
\newblock {\em Business \& Information Systems Engineering}, 6(2):97--110,
  2014.

\bibitem{wilde2020novel}
Henry Wilde, Vincent Knight, and Jonathan Gillard.
\newblock A novel initialisation based on hospital-resident assignment for the
  k-modes algorithm.
\newblock {\em arXiv preprint arXiv:2002.02701}, 2020.

\bibitem{vsegvic2014exploiting}
Sini{\v{s}}a {\v{S}}egvi{\'c}, Karla Brki{\'c}, Zoran Kalafati{\'c}, and Axel
  Pinz.
\newblock Exploiting temporal and spatial constraints in traffic sign detection
  from a moving vehicle.
\newblock {\em Machine vision and applications}, 25(3):649--665, 2014.

\bibitem{vsegvic2010computer}
S~{\v{S}}egvi{\'c}, K~Brki{\'c}, Z~Kalafati{\'c}, V~Stanisavljevi{\'c},
  M~{\v{S}}evrovi{\'c}, Damir Budimir, and I~Dadi{\'c}.
\newblock A computer vision assisted geoinformation inventory for traffic
  infrastructure.
\newblock In {\em 13th International IEEE Conference on Intelligent
  Transportation Systems}, pages 66--73. IEEE, 2010.

\bibitem{pallewatta2019microservices}
Samodha Pallewatta, Vassilis Kostakos, and Rajkumar Buyya.
\newblock Microservices-based iot application placement within heterogeneous
  and resource constrained fog computing environments.
\newblock In {\em Proceedings of the 12th IEEE/ACM International Conference on
  Utility and Cloud Computing}, pages 71--81, 2019.

\bibitem{zabrovskiy2018multi}
Anatoliy Zabrovskiy, Christian Feldmann, and Christian Timmerer.
\newblock Multi-codec dash dataset.
\newblock In {\em Proceedings of the 9th ACM Multimedia Systems Conference},
  pages 438--443. ACM, 2018.

\bibitem{opencv}
Framing a video.
\newblock
  \url{https://gist.github.com/SiNa88/c85d8cfac641918c6de8b4f31d8cdc22}.
\newblock [Online; accessed 12-March-2021].

\bibitem{ananthanarayanan2019video}
Ganesh Ananthanarayanan, Victor Bahl, Landon Cox, Alex Crown, Shadi Nogbahi,
  and Yuanchao Shu.
\newblock Video analytics-killer app for edge computing.
\newblock In {\em Proceedings of the 17th Annual International Conference on
  Mobile Systems, Applications, and Services}, pages 695--696, 2019.

\bibitem{stallkamp2012man}
Johannes Stallkamp, Marc Schlipsing, Jan Salmen, and Christian Igel.
\newblock Man vs. computer: Benchmarking machine learning algorithms for
  traffic sign recognition.
\newblock {\em Neural networks}, 32:323--332, 2012.

\bibitem{phoronix}
Phoronix test suite - benchmarking platform, and automated testing.
\newblock \url{https://www.phoronix-test-suite.com/}.
\newblock [Online; accessed 12-March-2021].

\bibitem{wilde2020matching}
Henry Wilde, Vincent Knight, and Jonathan Gillard.
\newblock Matching: A python library for solving matching games.
\newblock {\em Journal of Open Source Software}, 5(48):2169, 2020.

\bibitem{Li:2019}
Xiangbo Li, Mohsen~Amini Salehi, Yamini Joshi, Mahmoud~K. Darwich, Brad
  Landreneau, and Magdy~A. Bayoumi.
\newblock Performance analysis and modeling of video transcoding using
  heterogeneous cloud services.
\newblock {\em {IEEE} Trans. Parallel Distrib. Syst.}, 30(4):910--922, 2019.

\bibitem{cloudlet}
{The Alcatel-Lucent Ultimate Wireless Packet Core}.
\newblock
  \url{https://images.tmcnet.com/online-communities/ngc/pdfs/The-Alcatel-Lucent-Ultimate-Wireless-Packet-Core.pdf}.
\newblock [Online; accessed 12-March-2021].

\bibitem{wificloudlet}
Atlas 500 ai edge station (model: 3000).
\newblock
  \url{https://e.huawei.com/en/products/cloud-computing-dc/atlas/atlas-500}.
\newblock [Online; accessed 12-March-2021].

\bibitem{iperf3}
iperf - the ultimate speed test tool for tcp, udp and sctp.
\newblock \url{https://iperf.fr/}.
\newblock [Online; accessed 12-March-2021].

\bibitem{zhang2017towards}
Wuyang Zhang, Jiachen Chen, Yanyong Zhang, and Dipankar Raychaudhuri.
\newblock Towards efficient edge cloud augmentation for virtual reality mmogs.
\newblock In {\em Proceedings of the Second ACM/IEEE Symposium on Edge
  Computing}, pages 1--14, 2017.

\bibitem{topcuoglu2002performance}
Haluk Topcuoglu, Salim Hariri, and Min-you Wu.
\newblock Performance-effective and low-complexity task scheduling for
  heterogeneous computing.
\newblock {\em IEEE transactions on parallel and distributed systems},
  13(3):260--274, 2002.

\bibitem{c3}
The carinthian computing continuum.
\newblock \url{https://c3.itec.aau.at/}.

\bibitem{kimovski2021cloud}
Dragi Kimovski, Roland Math{\'a}, Josef Hammer, Narges Mehran, Hermann
  Hellwagner, and Radu Prodan.
\newblock Cloud, fog or edge: Where to compute?
\newblock {\em IEEE Internet Computing}, 2021.

\bibitem{exoscale}
European cloud hosting.
\newblock \url{https://www.exoscale.com/}.

\end{thebibliography}
\bibliographystyle{unsrt}
\end{document}